\documentclass[fleqn,usenatbib]{mnras}

\usepackage[T1]{fontenc}
\usepackage{ae,aecompl}

\usepackage{graphicx}	% Including figure files
\usepackage{amsmath}	% Advanced maths commands
\usepackage{amssymb}	% Extra maths symbols
\usepackage[normalem]{ulem}
\usepackage{xcolor}
\usepackage{newtxtext,newtxmath}

\def\ms{\mbox{$M_{\ast}$}}

\def\mha{\mbox{$M_{\rm HI}$}}
\def\HI{\mbox{$\rm HI$}}
\def\H2{\mbox{$\rm H_2$}}
\def\RHI{\mbox{$\rm R_{HI}$}}
\def\msun{\mbox{$M_{\odot}$}}

\def\ms{\mbox{$M_{\ast}$}}%Stellar Mass
\def\mha{\mbox{$M_{\rm H\,{\sc I}}$}}%HI mass
\def\mh{\mbox{$M_{h}$}}%Halo mass
\def\msun{\mbox{M$_{\odot}$}}%Solar mass

\def\HI{\mbox{H\,{\sc I}}}%Atomic hydrogen.
\def\H2{\mbox{$\rm H_{2}$}}%Molecular Hydrogen.

\def\RHI{\mbox{$R_{\rm H\,{\sc I}}$}}%HI mass-stellar mass ratio
\def\RH2{\mbox{$R_{\rm H_{2}}$}}%H2 mass-stellar mass ratio
\def\xG{\mbox{\texttt{xGASS}}}% xGASS
\def\mtr{\mbox{$M_{\rm tr}$}}%Hj conditional distribution LTGs

\def\gsmf{\mbox{GSMF}}

\def\mnras{\mbox{MNRAS}}
\def\apj{\mbox{ApJ}}
\def\aap{\mbox{A\&A}}
\def\apjl{\mbox{ApJ}}
\def\apjs{\mbox{ApJS}}
\def\aj{\mbox{AJ}}

\def\pasp{\mbox{PASP}}

\def\rmxaa{\mbox{RMxAA}}

\title[\HI\ distributions of centrals/satellites]{The \HI\ and stellar mass bivariate distribution of centrals and satellites for all, late- and early-type local galaxies}

\author[A. R. Calette et al.]{
A. R. Calette$^{1}$\thanks{E-mail: acalette@astro.unam.mx},
Vladimir Avila-Reese$^{1}$,
Aldo Rodr\'iguez-Puebla$^{1}$, \and 
Claudia del P. Lagos$^{2,3}$ and Barbara Catinella$^{2,3}$\\
$^{1}$Instituto de Astronom\'ia, Universidad Nacional Aut\'onoma de M\'exico, A. P. 70-264, 04510, Ciudad de M\'exico, M\'exico. \\
$^{2}$ International Centre for Radio Astronomy Research (ICRAR), M468, University of Western Australia, 35 Stirling Hwy,\\ Crawley, WA 6009, Australia. \\
$^{3}$ ARC Centre of Excellence for All Sky Astrophysics in 3 Dimensions (ASTRO 3D). \\
}

\date{Accepted XXX. Received YYY; in original form ZZZ}

\pubyear{2021}

\begin{document}
\label{firstpage}
\pagerange{\pageref{firstpage}--\pageref{lastpage}}
\maketitle

\begin{abstract}
We characterize the conditional distributions of the \HI\ gas-to-stellar mass ratio, $\RHI\equiv \mha/\ms$, given the stellar mass, \ms, of local galaxies from $\ms\sim 10^7$ to $10^{12}$ \msun\  separated into centrals and satellites as well as into late- and early-type galaxies (LTGs and ETGs, respectively). To do so, we use 1) the homogeneous  ``eXtended GALEX Arecibo SDSS Survey'', \xG \citep[][]{Catinella+2018}, by re-estimating their upper limits and taking into account them in our statistical analysis; and 2) the results from a large compilation of \HI\ data reported in \citet{Calette+2018}. We use the \RHI\ conditional distributions combined with the Galaxy Stellar Mass Function to infer the bivariate \mha\ and \ms\ distribution of all galaxies as well of the late/early-type and central/satellite subsamples and their combinations. Satellites are on average less \HI\ gas-rich than centrals at low and intermediate masses, with differences being larger for ETGs than LTGs; at $\ms>3-5\times 10^{10}$ \msun\ the differences are negligible. The differences in the \HI\ gas content are much larger between  LTGs and ETGs than between centrals and satellites. Our empirical \HI\ Mass Function is strongly dominated by central galaxies at all masses. The empirically constrained bivariate \mha\ and \ms\ distributions presented here can be used to compare and constrain theoretical predictions as well as to generate galaxy mock catalogues. 
\end{abstract}


%methods: statistical -- methods: observational -- methods: data analysis
%%%%%%%%%%%%%%%%%%%%%%%%%%%%%%%%%%%%%%%%%%%%%%%%%%

%%%%%%%%%%%%%%%%% BODY OF PAPER %%%%%%%%%%%%%%%%%%

\section{Introduction}
\label{Sec:Intro}

The evolution of galaxies depends on the interplay of many complex processes. Among them: gas cooling within dark matter haloes, transformation of the cool atomic hydrogen (\HI) gas into cold dense molecular hydrogen (\H2) clouds, the formation of stars in the densest regions of these clouds, and the ulterior feedback that stars and their explosions exert on the interstellar medium \citep[for a review, e.g.][]{Mo+2010book}. Therefore, the amounts of \HI\ and \H2\ gas relative to the stellar mass, morphological type, optical colours, and other galaxy properties, are crucial for understanding the evolutionary stage of local galaxies \citep[see e.g.,][]{Lagos+2011,Lagos+2014}. It is also well known that the environment, in particular whether a galaxy is central or satellite \citep[e.g.,][]{Kauffmann+2004,Boselli+2006,Davies+2019}, plays a role in the evolution of galaxies, so that information on the gas fractions of galaxies as a function of environment is also relevant \citep[e.g.,][and more references therein]{Brown+2017,Stevens+2019}. 

Although \HI\ gas is the dominant component in the interstellar medium of local galaxies \citep[][]{Fukugita+1998}, its detection is not easy because of its weak 21-cm emission line. Great efforts have been made to build large radio \HI\ surveys as the \HI\ Parkes All-Sky Survey \citep[HIPASS;][] {Meyer+2004} and Arecibo Fast Legacy ALFA Survey \citep[ALFALFA;][]{Giovanelli+2005,Haynes+2011}. However, these blind radio surveys are not yet as deep and do not cover such large areas as the optical/infrared extragalactic surveys, and are affected by strong selection effects. Thus, the inferred  \HI\ gas scaling relations, \HI\ velocity function, as well as other correlations and \HI\ spatial distributions, result biased if based on detections only \citep[c.f.][]{Meyer+2007,Haynes+2011,Huang+2012,Papastergis+2013,Maddox+2015,Guo+2017,Calette+2018}. So, volume corrections or strategies like \HI\ spectral stacking \citep[e.g.,][]{Brown+2015} are required to infer approximations to the intrinsic relations and distribution functions. Another way to attempt to overcome the strong selection effects of blind \HI\ radio surveys is to construct ``well controlled'' \HI\ samples by means of radio follow-up observations of optically selected galaxy samples \citep[e.g.,][]{Wei+2010a,Catinella+2013,Catinella+2018,Papastergis+2012,Kannappan+2013,Boselli+2014,Eckert+2015,Stark+2016,vanDriel+2016,Masters+2019}. These samples were designed for a variety of scientific goals, and as a result they are diverse and heterogeneous, covering different stellar mass ranges, distances, and \HI\ flux detection limits, and commonly they are far from complete in stellar mass.

\subsection{The HI conditional distributions of late and early-type galaxies}

In \citet[][and with updates in \citealp{Rodriguez-Puebla+2020}, hereafter Papers I and II respectively]{Calette+2018}, we undertook the task of compiling and homogenizing from the literature many \HI\ galaxy samples such the ones listed above (including most of them), with the additional requirement of information on galaxy morphology being available. The latter was done as the \HI\ gas content of galaxies strongly depends on morphology, hence it is more appropriate to analyze it separately for galaxies of at least two broad morphological groups. 
We took into account the reported upper limits for the radio non-detections, and after homogenizing them to a distance of $\sim 25$ Mpc and similar signal-to-noise ratio detection limit %and correcting some of them for distance limitations, 
we applied a survival analysis to determine gas correlations. As a result, we were able to constrain not only the mean \mha--\ms\ relation for late- and early-type galaxies (LTG and ETG, respectively) down to $\ms\sim 10^7$ \msun, but the respective conditional probability density distribution functions (PDFs) of \mha{} given \ms, $P(\mha|\ms$). From these PDFs, one can calculate any moment of the distributions, in particular the standard deviation around the mean relation, as well as the percentiles. 

In Paper II we used the well-constrained Galaxy Stellar Mass Function (\gsmf) for all, late- and early-type galaxies down to $\sim 10^7$ \msun{} computed there, and combined them with the $P(\mha|\ms)$ distributions to generate the bivariate (joint) \ms\ and \mha\ distribution function. By projecting this bivariate distribution into the \HI\ axis, we obtained the \HI\ MFs, for LTGs and ETGs, as well as for all galaxies. We have shown that our empirical \HI{} MF (corresponding to a volume-limited sample complete above $\mha\sim 10^8$ \msun)
agrees well with those measured from blind radio surveys.

%%%%%%%%%%%%%%%%FIGURE%%%%%%%%%%%%%%%
\begin{figure*}
	%trim=l b r t
	\includegraphics[trim = 0mm 30mm 0mm 23mm, clip, width=\textwidth]{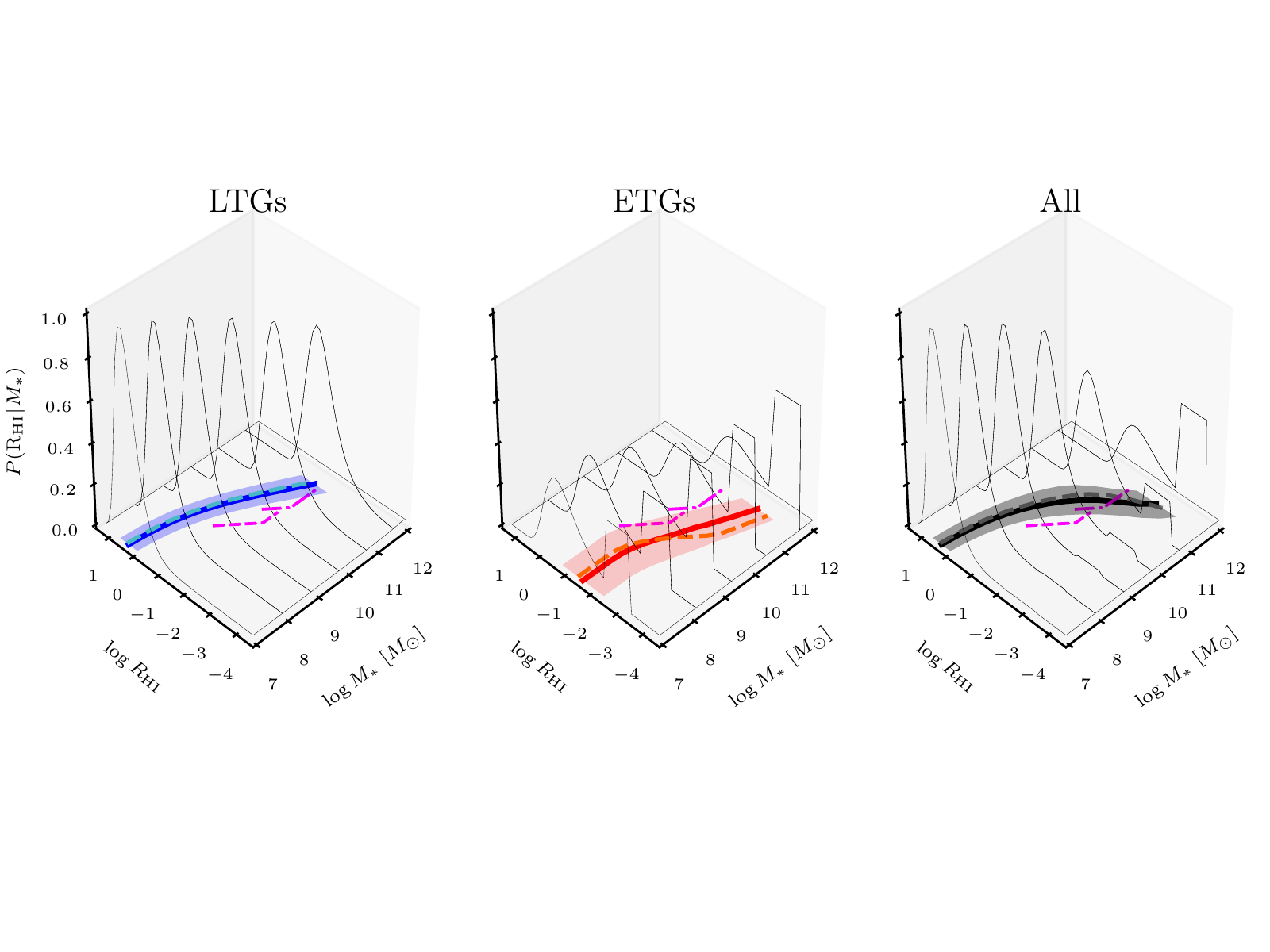} %,logRHI-logMs-PDFs-total-LTGs-ETGs.pdf height=0.70\textwidth, height=180pt
	%\textit{Left panels:}
		\caption{The \RHI\ gas conditional PDFs of LTGs, ETGs, and all galaxies as a function of \ms\ from \citet[][]{Calette+2018} and updated in \citet[][]{Rodriguez-Puebla+2020}. In the projected $\log\RHI$-$\log\ms$ planes, the logarithmic mean relations and their standard deviation are shown with thick solid lines and shaded areas, respectively. The dashed lines correspond to the medians instead of the logarithmic means. The magenta dashed and dot-dashed lines show the \xG\ detection limits, see Section \ref{observational-data}.
	} 
	\label{fig:logRHI-logMs-PDFs} 
\end{figure*}

%%%%%%%%%%%%%% END FIGURE%%%%%%%%%%%%%

In Paper I we showed that the conditional PDFs of the \HI-to-stellar mass ratio, $\RHI\equiv\mha/\ms$, given \ms\ can be well described by a Schechter-type function for LTGs \citep[see also][]{Lemonias+2013} and a (broken) Schechter-type function plus a top-hat function for ETGs, having the latter significantly lower values of \HI{} gas content than the former at fixed stellar mass. 
In Figure \ref{fig:logRHI-logMs-PDFs} we reproduce these $P(\RHI|\ms)$ distributions 
as a function of \ms, left-hand and medium panels, along with the respective logarithmic mean \RHI--\ms\ relations and standard deviations (first and second moments of the $\log\RHI$ distributions), solid lines and shaded regions, respectively. The thick dashed lines are the corresponding medians.
While for LTGs, both the mean and median \RHI--\ms\ relations are similar, for ETGs, they differ, specially at the high-mass side. 
The right-hand panel shows the resulting \RHI\ conditional PDFs for all galaxies as well as the respective first and second moments.  
We infer the \RHI\ conditional distribution for all galaxies by using the fractions of ETGs and LTGs as a function of \ms\ from the Sloan Digital Sky Survey (SDSS) based on the \citet[][]{Huertas-Company+2011} morphological classification, corrected for volume completeness (see Paper II for details).  

\subsection{The HI gas content of central and satellite galaxies}

The $P(\RHI|\ms)$ distributions and the main relations shown in Fig. \ref{fig:logRHI-logMs-PDFs} do not distinguish between central and satellite galaxies.  Though it is not clear whether the \HI\ gas fraction of galaxies correlates directly or not with the large-scale environment (see Paper I for a discussion, and the references therein), at the level of central and satellite galaxies, there are differences with latter having lower \HI\ gas contents at a given stellar mass than the former \citep[e.g.,][but see \citealp{Lu+2020}]{Stark+2016,Brown+2016}. 

The goal of this paper is to introduce adequate functions to our empirical \HI\ conditional PDFs for LTGs and ETGs in such a way that they can be separated into central and satellite galaxies. For this, we will use the recent \HI\ observational survey eXtended GALEX Arecibo SDSS Survey \citep[\texttt{xGASS};][]{Catinella+2018}. \texttt{xGASS} is an homogeneously constructed \HI, ultraviolet (UV), and optical galaxy sample with well-defined limits in \RHI, \ms, and volume. Since this survey was constructed from SDSS, most of the galaxies can be separated into centrals and satellites making use of the \citet[][]{Yang+2007,Yang+2012} halo-based group definition applied to SDSS. 
Thus, from \texttt{xGASS} we calculate the ratios of central and satellite to total \RHI\ conditional PDFs as a function of \ms\ for both late- and early-type galaxies. These ratios are applied to our empirical LTG and ETG \RHI\ PDFs to separate them into centrals and satellites.

This paper is organized as follows. In Section \ref{observational-data} we describe the \xG\ survey and our processing, in particular for the upper limits.  Section 3 presents the results of our statistical analysis of \xG: the \HI-to-stellar mass relations for LTGs and ETGs separated into centrals and satellites, as well as the respective \HI\ conditional PDFs and join fits of analytic functions to these. In Section 4 we use the \xG\ \HI\ conditional PDFs to separate the distributions constrained in Papers I and II into centrals and satellites. By combining these distribution with the GSMF, we construct the full bivariate (joint) \ms\ and \mha\ distributions of all galaxies as well as of subsamples of centrals/satellites, LTGs/ETGs, and their combinations. Section 5 is devoted to discussing the caveats and implications  of our results. Finally, in Section 6 we present a summary of the paper and the  conclusions.

\section{Analysis of the \xG\ survey}
\label{observational-data}

The survey \texttt{xGASS} \citep[][]{Catinella+2018} is an \RHI-limited census of 1179 galaxies selected by redshift and \ms\ in the ranges $0.01\leq z \leq 0.05$ and $10^{9}\msun\leq\ms\leq 10^{11.5}\msun$, respectively. The sample galaxies were drawn from the intersection of SDSS DR7 \citep{Abazajian+2009}, GALEX \citep{Martin+2005} and projected ALFALFA footprints \citep{Haynes+2011}.  
The \texttt{xGASS} consists of two samples: (1) 
\texttt{GASS} \citep{Catinella+2010,Catinella+2012,Catinella+2013}, a sample of galaxies with $\ms>10^{10}$ \msun\ and redshift $0.025\leq z \leq 0.05$, and (2) the low-stellar mass extension of \texttt{GASS} \citep[hereafter low-\texttt{GASS}][]{Catinella+2018}, a sample of galaxies with stellar masses in the range $10^{9}\msun\leq\ms\leq 10^{10.2}\msun$ and redshift $0.01\leq z \leq 0.02$. Both samples were constructed in such a way that the stellar mass distribution of the targets is roughly flat. The \xG\ survey is the most complete \HI\ observational study of a local optically based representative galaxy sample to date. 

In \xG, the \HI\ mass is obtained from the \HI\ observations of ALFALFA $\alpha.40$ or the Cornell \HI\ digital archive \citep{Springob+2005}.
For galaxies with no \HI\ information, observations were performed using the Arecibo Radio Telescope with the strategy of observing the targets until detected or until a limit of a few percent in \RHI\ ratio is reached. The detection limits for each sample are:
\begin{itemize}
\item \texttt{GASS}:  $\RHI>0.015$ for galaxies with $\ms>10^{10.5}$\msun\ and a constant \HI\ mass limit of $\mha=10^{8.7}$ \msun\ for galaxies with lower stellar masses.

\item low-\texttt{GASS}: $\RHI>0.02$ for galaxies with $\ms>10^{9.7}$\msun\ and a constant \HI\ mass limit of $\mha=10^{ 8}$ \msun\ for lower mass galaxies.
\end{itemize}
The detection limits in \RHI\ considered mainly  the telescope sensitivity, integration time, and the redshift range of the surveys.

\subsection{Morphology and central/satellite designations for \texttt{xGASS} galaxies}
\label{subsec:morph}

At fixed \ms, the gas content in galaxies varies significantly with morphology \citep[e.g.,][]{Kannappan+2013, Boselli+2014b,Calette+2018}.
Thus, we introduce a morphological characterization for \xG\ galaxies that complements the dependence on stellar mass. 
Here, we use the \cite{Huertas-Company+2011} automated morphological classification for $\sim$700 000 galaxies from the SDSS DR7 spectroscopic sample, where each galaxy has a probability of being elliptical, S0, Sab and Scd by means of support vector machines (SVM) method and the \cite{Fukugita+2007} sample as a training set.\footnote{In \S\S \ref{sec:DS18} we discuss how our our results do change when applying an alternative morphological classification scheme.}
On the other hand, \cite{Meert+2015} calibrated  \cite{Huertas-Company+2011} probabilities to T-types using a simple linear model given by,
\begin{equation}\label{eq:Ttypes}
    T=-4.6\cdot P({\rm Ell})-2.4\cdot P({\rm S0})+2.5\cdot P({\rm Sab})+6.1\cdot P({\rm Scd})
\end{equation}
The latter was constrained using the visual classification of \cite{Nair+2010} by a linear regression. Using eq. (\ref{eq:Ttypes}) and the probability classification from \cite{Huertas-Company+2011}, we assign T-types to \texttt{xGASS} galaxies. Of the 1179 galaxies in the \xG\ sample we find that 1150 are in the \cite{Huertas-Company+2011} morphology catalogue and only consider these for our analysis. 

We separate \xG\ galaxies into two broad morphological groups: LTGs and ETGs.
We consider ETGs as those galaxies with $T<0.5$ and LTGs as those with $T\ge 0.5$ following \cite{Meert+2015}. The above corresponds respectively to S0 or earlier and Sab or later morphologies, see their eq. (8) for details.

To segregate galaxies into centrals and satellites we use the \texttt{xGASS} flag \texttt{env\_code\_B} defined as\footnote{xGASS data description: \url{https://xgass.icrar.org/assets/data/xGASS_representative_sample.readme}}:
    \[ \mbox{\texttt{env\_code\_B} } = \begin{cases} 
          0: & \mbox{satellite} \\
          1: & \mbox{isolated central} \\
          2: & \mbox{group central} \\
          -1: & \mbox{not in group catalogue} 
       \end{cases}
    \]
We consider centrals those galaxies with \texttt{env\_code\_B}=1 or 2. The term isolated central does not imply what typically is known in the literature as an isolated environment but it refers to the presence of only one galaxy within the halo. Satellites are those with \texttt{env\_code\_B}=0. As described in \citet[][]{Janowiecki+2017}, for determining whether a galaxy is central or satellite in \xG, the authors used the \citet[][]{Yang+2007} halo-based group catalogue updated to the SDSS DR7. For \xG, the ``modelB'' group catalogue was adopted, and cases of ``galaxy shredding'' and false pairs have been resolved by visual inspection \citep[see details in][]{Janowiecki+2017}.
Fortunately, only a small fraction of \xG\ galaxies, $2\%$, are not in the \citet[][]{Yang+2007} ``modelB'' catalogue or suffer from galaxy shredding and false pairs. Approximately 30\% of \xG\ galaxies are classified as satellites in groups, $\sim 50\%$ as isolated centrals, and $\sim 20\%$ as centrals (the most massive member) in groups.  
The central/satellite designation adopted for the \xG\ survey has been used in several works for studying the effects of environment on the gas content of galaxies \citep[e.g.,][]{Janowiecki+2017,Janowiecki+2020,Stevens+2019,Cortese+2020,Watts+2020}. Nevertheless, it should be stressed that galaxy group finders like the \citet[][]{Yang+2005,Yang+2007} halo-based finder may suffer from membership allocation and central/satellite designation errors. In \S\S \ref{sec:explorations}, we discuss this caveat and how it can affect the results obtained in this paper. 

%%%%%%%%%%%%%%%%FIGURE%%%%%%%%%%%%%%%
\begin{figure*}
	%trim=l b r t
	\includegraphics[trim = 6mm 65mm 90mm 13mm, clip, width=0.8\textwidth]{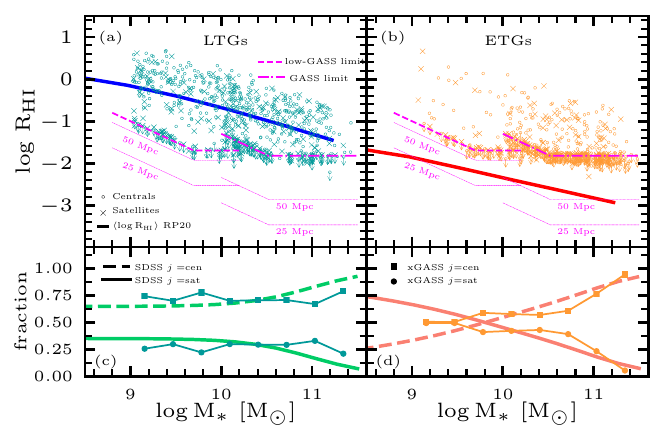} %, height=0.70\textwidth, height=180pt 240pt height=300pt xGASS_logRHI_xGASS.pdf
	\caption{Presentation of the \xG\ sample. \textit{Panel (a):} LTGs in the $\log\RHI$-$\log\ms$ diagram, with centrals and satellites  plotted as empty circles and crosses, respectively. The downward arrows indicate the reported upper limits for non detected galaxies in radio. Dot-dashed and dashed lines show the imposed limit detection in the \texttt{GASS} and the low-\texttt{GASS} samples, respectively. We reproduce the logarithmic mean of LTGs obtained in Papers I and II with blue solid line  (see also Fig. \ref{fig:logRHI-logMs-PDFs}). \textit{Panel (b):} Same as panel (a) but for ETGs. \textit{Panel (c):} Fraction of LTGs that are satellites (circles) or centrals (squares) as a function of \ms. The respective fractions as inferred from SDSS DR7 based on the \citet{Huertas-Company+2011} morphological classification and the \citet[][]{Yang+2012} central/satellite division are plotted with the dashed and solid lines, respectively. \textit{Panel (d):} Same as panel (c) but for ETGs.
	} 
	\label{fig:xGASS-data-and-fracs} 
\end{figure*}

%%%%%%%%%%%%%% END FIGURE%%%%%%%%%%%%%

\subsection{Final {\rm \xG} sample and selection effects}

The final sample of \xG\ galaxies with morphology and central/satellite classifications includes 1134 objects. 
In panels (a) and  (d) of Figure \ref{fig:xGASS-data-and-fracs}, we present these galaxies in the \RHI--\ms\ plane separated into LTGs and ETGs, respectively. In each panel central and satellite galaxies are plotted with open circles and crosses, respectively, and upper limits are shown with downward arrows. The number of galaxies with upper limits is significant, 55\% for ETGs and  17\% for LTGs. 
The dot-dashed and dashed lines show the imposed detection limit in the \texttt{GASS} and \texttt{GASS}-low samples, respectively. Most of the upper limits pile up close to these lines. However, since galaxies are at different distances the distribution of the upper limits is somewhat scattered.
In the same panels, we reproduce the logarithmic means of LTGs and ETGs obtained in Papers I and II. Their corresponding \RHI\ conditional distributions at different stellar masses are shown respectively in the panels (b) and (e). In these panels, we also reproduce the \xG\ detection limits.
Clearly the empirical distribution of \RHI\ is truncated by the \xG\ detection limits. This truncation is abrupt for ETGs, which are even above the first moments of the empirical \RHI\ PDFs (the red solid line in panel d). 

An upper limit in \HI\ mass is reported when a galaxy in a given survey has not been detected in the 21-cm line for the defined integration time and above a given signal-to-noise ratio. The \HI\ mass upper limit is calculated using the respective \HI\ flux detection limit and the distance to the galaxy, $M^{\rm u.l.}_{\rm HI}\propto D(z)^2$.
When inferring any correlation or probability distribution from \mha, it is mandatory to account for upper limits.
In \S\S \ref{KM-estimator} we describe the survival analysis we follow to do so. In addition, it is important to note that the \texttt{xGASS} upper limits are high and notably truncate the low-side \RHI\ distribution, specially for ETGs. This is due to the large distances in this survey, in particular for \texttt{GASS}.  In fact, in galaxy samples at closer distances than \xG, a fraction of their galaxies were detected in \HI\ with \RHI\ values below the \texttt{xGASS} detection limits, for instance, in ATLAS$^{3D}$ \citep[][]{Cappellari+2011, Serra+2012} and Herschel Reference Survey \citep[HRS;][]{Boselli+2010,Boselli+2014}. On the other hand, the \HI\ detection limits of these closer galaxy samples, after taking into account the differences in the observational and instrumental settings, result in much lower upper limits than those from \texttt{xGASS}, in particular for the \texttt{GASS} sample. Thus, the upper limits from \texttt{xGASS} are biased high due to distance selection effect. 
Following Paper I and based on some assumptions, in \S\S \ref{sec:upper-limits} we attempt to correct for this bias in the upper limits. 

Panels (c) and (f) of Figure \ref{fig:xGASS-data-and-fracs} present respectively the \texttt{xGASS} fraction of LTGs and ETGs that are satellites, circles, or centrals, squares, as a function of \ms.  In the same panels, the solid lines correspond to fit to the satellite fractions for LTGs and ETGs from the \citet{Yang+2012} SDSS DR7 galaxy group catalogue (the dashed lines are the respective central fractions and they are by definition the complements of the solid lines; see Appendix \ref{app:fractions}). 
At this point, it is important to ask ourselves if \xG\ suffers of selection effects that could bias the sample by morphology (for the morphological classification adopted here, i.e., \citealp{Huertas-Company+2011}) and by environment. A bias in the morphology is not relevant when the inferred \RHI--\ms\ relations and \RHI\ distributions are determined separately for LTGs or ETGs. However, this possible bias is expected to affect the relations and distributions for all, central, and satellite galaxies when averaging among LTGs and ETGs. 

In Figure  \ref{fig:fractions} in Appendix \ref{app:fractions}, we compare the ETG and satellite fractions as a function of \ms\ from \xG\ with those measured from SDSS DR7 (panels (a) and (d), respectively). As seen, the \xG\ fraction of satellites as a function of \ms\ roughly agree with that from the whole SDSS DR7 (the fraction of centrals is the complement). 
However, this is not the case for the fraction of ETGs (the fraction of LTGs is the complement): \xG\ selects systematically a higher fraction of ETGs than SDSS up to $\ms\sim 10^{11}$ \msun. Obviously, the differences remain when considering only central or satellite galaxies,  but they are larger for satellites, compare panels (b) and (c). For $\ms\gtrsim 10^{11}$ \msun, the difference inverts. 
Note that the flat distribution in mass of \xG\ is not an issue in Figure  \ref{fig:xGASS-data-and-fracs} given that the comparisons between fractions are at a given \ms. 

For the inferences in Section \ref{results} of the \RHI--\ms\ relations and \RHI\ distributions given \ms\ corresponding to all galaxies (LTGs + ETGs), to all centrals (LTGs + ETGs) and to all satellites (LTGs + ETGs), we introduce weights for the \xG\ galaxies in order to be consistent with the fractions of ETGs as a function of \ms\ for both the samples of centrals and satellites from the SDSS DR7 (panels b and c of Fig. \ref{fig:fractions}).  The weighting procedure is described in Appendix \ref{app:fractions}. 

\subsection{Reestimating the HI upper limits}
\label{sec:upper-limits}

As mentioned above, when comparing the distribution of  \texttt{xGASS} galaxies in Figure \ref{fig:xGASS-data-and-fracs} with the respective empirical \HI\ conditional PDFs, shown in Figure \ref{fig:logRHI-logMs-PDFs}, we note that the \texttt{xGASS} detection limits truncate significantly the \HI\ conditional PDFs of ETGs (the corresponding \RHI--\ms\ relation lies even below the detection limits). In contrast, for LTGs the truncation is not significant given the high \HI\ gas contents for most of these galaxies.  
We ask ourselves: where would non-detected ETGs in \texttt{xGASS} appear in the \RHI--\ms\ plane if they were observed with the same instrument, observational setup, and allowed signal-to-noise ratio but at lower distances?
The thin dash-dotted and dotted lines in Figure \ref{fig:xGASS-data-and-fracs}, labeled respectively as 50 Mpc and 25 Mpc, show the shift that the \texttt{GASS} and \texttt{GASS}-low detection limits would have at these distances.\footnote{Notice that the \texttt{GASS} and the \texttt{GASS}-low samples are at a median distance  of 165 and 65 Mpc, respectively.}  
We see that at a distance of $\sim25$ Mpc, the detection limits lie now below the \RHI--\ms\ relation of ETGs.
Fortunately, there are close samples of ETGs with radio observations. As mentioned above, this is the case of the  ATLAS$^{\rm 3D}$ survey (median distance of 25 Mpc), which has been used  in Paper I for reestimating the \mha\ upper limits of  \texttt{GASS} ETGs, and eventually, for assigning detection values to a fraction of these upper limits.   

Here, we follow a procedure similar as in Paper I for reestimating the \mha\ upper limits of \xG. 
We emphasize that the procedure in Paper I is based on the assumption that the \HI\ gas content at a fixed \ms\ of galaxies at distances $\sim 25$ Mpc (the median distance of ATLAS$^{\rm 3D}$) is statistically the same as that of galaxies up to 100-200 Mpc (the distances of \texttt{GASS} galaxies). Under this assumption, the \HI\ observations for ATLAS$^{\rm 3D}$ (and also HRS) galaxies allowed us to re-estimate {\it in a statistical sense} the \RHI\ upper limits of \texttt{GASS} galaxies and to assign (detected)  \mha\ values to a fraction of them. Of course, only future deeper radio observations for each galaxy could provide a measure of its true \HI\ mass or a new improved upper limit. Performing a similar analysis to \texttt{GASS}-low  will require information of a survey such as ATLAS$^{\rm 3D}$. Unfortunately, this survey extends only down to stellar masses slightly smaller than $\sim10^{10}$ \msun, making the extension to \texttt{GASS}-low impractical at this point.

In the case of LTGs, most of them are detected in \texttt{GASS} despite their relatively shallow \HI\ detection limit.  On the other hand, for LTGs there is not a closer and homogeneous sample similar to ATLAS$^{\rm 3D}$. Thus, in Paper I, we did not attempt to correct the upper limits of LTGs from \texttt{GASS} by the distance effect. 
For \texttt{GASS}-low, the fraction of radio-detected LTGs from closer samples below the \texttt{GASS}-low  detection limit is slightly larger than in \texttt{GASS}. The overall fraction of upper limits for LTGs in \texttt{xGASS} is 17\%. Therefore, following the above argument for ETGs,
it would be desirable to attempt to also re-estimate the upper limits of LTGs.

As mentioned above, there are not close samples, as ATLAS$^{\rm 3D}$, with more or less well defined detection limits in \RHI\ for $\ms<10^{10}$ \msun, both for early- and late-type galaxies. However, we can use the empirically constrained \RHI\ distributions in Papers I and II  to re-estimate the reported  \texttt{GASS}-low upper limits due to their bias by distance. Even more, to homogenize our procedure 
we decided here to use these empirical  
distributions for both \texttt{xGASS} ETGs and LTGs. For \texttt{GASS} ETGs, the re-estimation of upper limits obtained here are very similar to those in Paper I. 
Following the discussion above, in Appendix \ref{App:upper-limits} we describe in detail our procedure to re-estimate the upper limits of ETGs and LTGs for \xG. 

\subsection{Statistical analysis including HI upper limits}
\label{KM-estimator}

In order to estimate from \xG\ the \RHI--\ms\ relations separated into central and satellite galaxies or, even more, the full \RHI\ conditional PDFs given \ms, as in Paper I, the upper limits should be taken into account adequately. 
In observational Astrophysics, we are often interested on particular astronomical objects (e.g. stars, galaxies, etc) and in order to design samples to study them, we set a selection criteria based on a given property, $P_{1}$, to construct such observational samples (for example stellar mass or luminosity). But there are situations when we are also interested in another property, $P_{2}$ (for example \HI\ content). Nevertheless, due to instrumental limitations we cannot always measure the property $P_{2}$ in all objects, instead we assign upper limits or ``censored" data values. In such situation it is necessary to build a parent sample based on a well studied property $P_{1}$ and then examine for the property of interest $P_{2}$ from property $P_{1}$. 
The above description is exactly the case for the \xG\ sample, in which $P_{1}=\ms$ and  $P_{2}=\mha$.

To use both detections and upper limits from \xG, in this work we rely on Kaplan-Meier (KM) non-parametric estimator \citep{Kaplan+1958} specifically developed for the analysis of censored data in clinical research, but properly adapted to astronomical data by \citet{Feigelson+1985}. For a given sample, the KM estimator allows us to obtain the
cumulative distribution function (CDF) when including censored data and from different statistical estimators can be calculated. However, to obtain reliable results, it is recommended that the fraction of censored data (upper limits) be less than $\sim 50\%$.
We construct the \RHI\ CDFs at different stellar mass bins. After applying the corrections to the ETG upper limits (see \S\S \ref{sec:upper-limits}), the minimum \RHI\ values (censored data) used in the KM estimator are around $-3.0<\log\RHI<-3.5$, and the CDFs at these values start with fractions typically of 0.3--0.4. This means that around 30-40\% of ETGs have \RHI\ upper limits. As mentioned in the footnote of Appendix \ref{ETG-corrs}, in Paper I we assigned real values (detections) to these galaxies by assuming they follow a top-hat function of width $\sim 1$ dex below the minimum upper limit value of the given mass bin. Our main argument was that even quiescent ETGs should have \HI\ gas fractions larger than a few $10^{-5}$, taking into account stellar mass loss and some eventual gas capture from minor mergers and cosmic accretion.

%%%%%%%%%%%%%%%%FIGURE%%%%%%%%%%%%%%%
\begin{figure*}
	%trim=l b r t
	\includegraphics[trim = 5mm 43mm 55mm 14mm, clip, width=\textwidth]{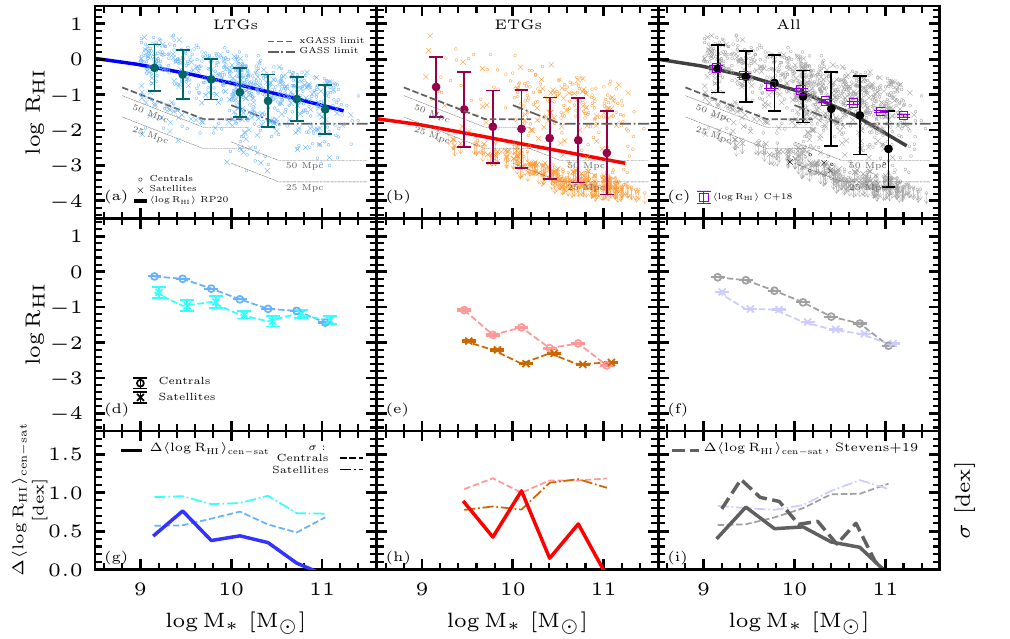} %, height=0.70\textwidth, height=180pt
	\caption{
	\textit{Upper panels:} \xG\ galaxies in the $\log\RHI-\log\ms$ diagram, as in Figure \ref{fig:xGASS-data-and-fracs} but after applying corrections to the upper limits (see text). The symbols with error bars are the logarithmic means and standard deviations in \ms\ bins obtained with the KM estimator for taking into account upper limits (the data are presented in tabulated form in the Supplementary Material). The solid lines show the mean \RHI--\ms\ relations from Paper II. In panel (c), the violet empty squares are the logarithmic means as reported in \citet[][]{Catinella+2018}.
	\textit{Middle panels:} Logarithmic means and their error on the mean in \ms\ bins obtained with the KM estimator for the subpopulations of central (open circles with error bars) and satellite (crosses with error bars) galaxies, for late-type, early-type, and all galaxies from left to right. 
	\textit{Lower panels:} Second moments of the $\log\RHI$ distributions from the KM estimator for the subpopulations of central and satellite galaxies (dashed lines) showed in the upper panels. The solid lines are the relative differences between the means of central and satellite subpopulations showed in the medium panels. The long-dashed line in panel (i) corresponds to the relative differences between the medians of centrals and satellites as reported for \xG\ in \citet{Stevens+2019}. 
	}
	\label{fig:xGASS-correlations} 
\end{figure*}
%%%%%%%%%%%%%% END FIGURE%%%%%%%%%%%%%

\section{Results from \xG}
\label{results}

\subsection{Correlations for all, central, and satellite galaxies}

In the upper panels of Fig. \ref{fig:xGASS-correlations} we plot again the \xG\ data as in Fig. \ref{fig:xGASS-data-and-fracs} but after applying to the upper limits the procedure described in \S\S \ref{sec:upper-limits}; we added a third panel showing the whole sample (LTGs+ ETGs). 
For each \ms\ bin of width $\Delta\log\ms=0.31$ dex, we use the procedure based on the KM estimator described in \S\S \ref{KM-estimator} to calculate the mean logarithmic value of \RHI\ and the standard deviation at each \ms\ bin. The results are plotted with circles and error bars. For comparison, the thick solid line in each panel is the respective logarithmic  mean relation as obtained in Paper II and also reproduced in Figs. \ref{fig:logRHI-logMs-PDFs} and \ref{fig:xGASS-data-and-fracs} above. 
For LTGs, \xG\ is in very good agreement with our empirical relation from Paper II. In the case of ETGs, the averages of \xG\ galaxies (after re-scaling the upper limits by the distance bias) are slightly above than the corresponding relation from Paper II but within the standard deviations. Note that these upper limits lie now around the \texttt{GASS} and low-\texttt{GASS} detection limits shifted to a distance of 25 Mpc. 
 
 In the right-hand panel of Figure \ref{fig:xGASS-correlations}, corresponding to all galaxies, we reproduce the logarithmic mean \RHI\ values reported by \citet[][]{Catinella+2018}, violet squares. These authors calculated the means (i) setting the \HI\ mass of non-detections to their upper limit values (this leads to overestimate the mean), and (ii) applying weights to correct for the stellar mass bias of the sample, that is, to make the sample mass complete in volume.  Regarding  (ii), it is not expected to be relevant for the means calculated in small mass bins since the weights are roughly the same for similar masses. 
 At low and intermediate stellar masses our means are in good agreement with those from \citet[][]{Catinella+2018} but at the highest masses, where ETGs dominate, our means are lower than those reported by these authors. This is due to the special treatment we applied to adequately include the upper limits of ETGs. 
 Recall that we also weighted \xG\ galaxies by morphology and environment to agree with the SDSS DR7 fractions of ETGs and satellites as a function of \ms, see \S\S \ref{subsec:morph}. The weights correct mainly the excess of ETGs in \xG\ with respect to SDSS up to $\ms\sim 10^{11}$ \msun\ and the lack at larger masses (the latter specially applies for satellites), see Figure \ref{fig:fractions}. Therefore, the average values plotted in Figure \ref{fig:xGASS-correlations} for all galaxies are weighted towards LTGs up to $\ms\sim 10^{11}$ \msun\ and against them at higher masses. 
 
In Appendix \ref{App:no-corrections} we present results for \xG\ without taking into account our procedure for the upper limits, nor the correction by morphology/environment. For LTGs, the results are almost indistinguishable from those presented here but for ETGs, for which the fraction of non-detections is high, for $\ms>5\times 10^9$ \msun, the mean \RHI\ values and their standard deviations obtained with the KM estimator are very uncertain and can be taken just as an upper bound. For the whole sample, combining LTGs and ETGs, we show that the weights by morphology slightly increase the mean \RHI\ values for masses below $\ms\sim 5\times 10^{10}$ \msun, while for the highest masses, the weights decrease the mean \RHI\ by $\sim 0.3$ dex.

The middle panels of Figure \ref{fig:xGASS-correlations} show $\langle\log\RHI\rangle$ and the errors of the mean, this time for central and satellite galaxies, separately. The solid lines connect the respective means showed in the upper panels.
Centrals have on average slightly higher \HI\ gas fractions than the average. For satellites, the differences are more pronounced especially towards lower stellar masses.
Overall, centrals have higher \HI\ gas contents than satellites, in particular at lower masses. 

In the lower panels of Fig. \ref{fig:xGASS-correlations}, we plot the logarithmic standard deviations for centrals and satellites at each mass bin for LTG, ETGs, and all galaxies. The population of ETGs presents larger scatter around the \RHI--\ms\ relations for centrals and satellites than LTGs. 
In each of the lower panels of Figure \ref{fig:xGASS-correlations}, we plot also the relative differences between the corresponding central and satellite means, $\Delta\langle\log\RHI\rangle_{\rm cen-sat}\equiv \langle\log\RHI_{\rm ,cen}\rangle - \langle\log\RHI_{\rm ,sat}\rangle$ (thick solid lines), plotted in the medium panels. As seen, these differences tend to be smaller than the corresponding standard deviations, both for LTGs and ETGs, specially at larger masses.
On average, satellite galaxies have lower \HI\ gas contents than centrals, specially at low masses.
Finally, in panel (i), corresponding to all galaxies, we reproduce the relative differences between the central and satellite medians reported in \citet{Stevens+2019} for \xG\ (long-dashed line). Despite them measuring medians and us logarithmic means and 
them setting non-detections to their upper limit values, the agreement is reasonable.

%%%%%%%%%%%%%%%%FIGURE%%%%%%%%%%%%%%%
\begin{figure*}
	%trim=l b r t
	\includegraphics[trim = 4mm 40mm 0mm 10mm, clip, width=0.9\textwidth]{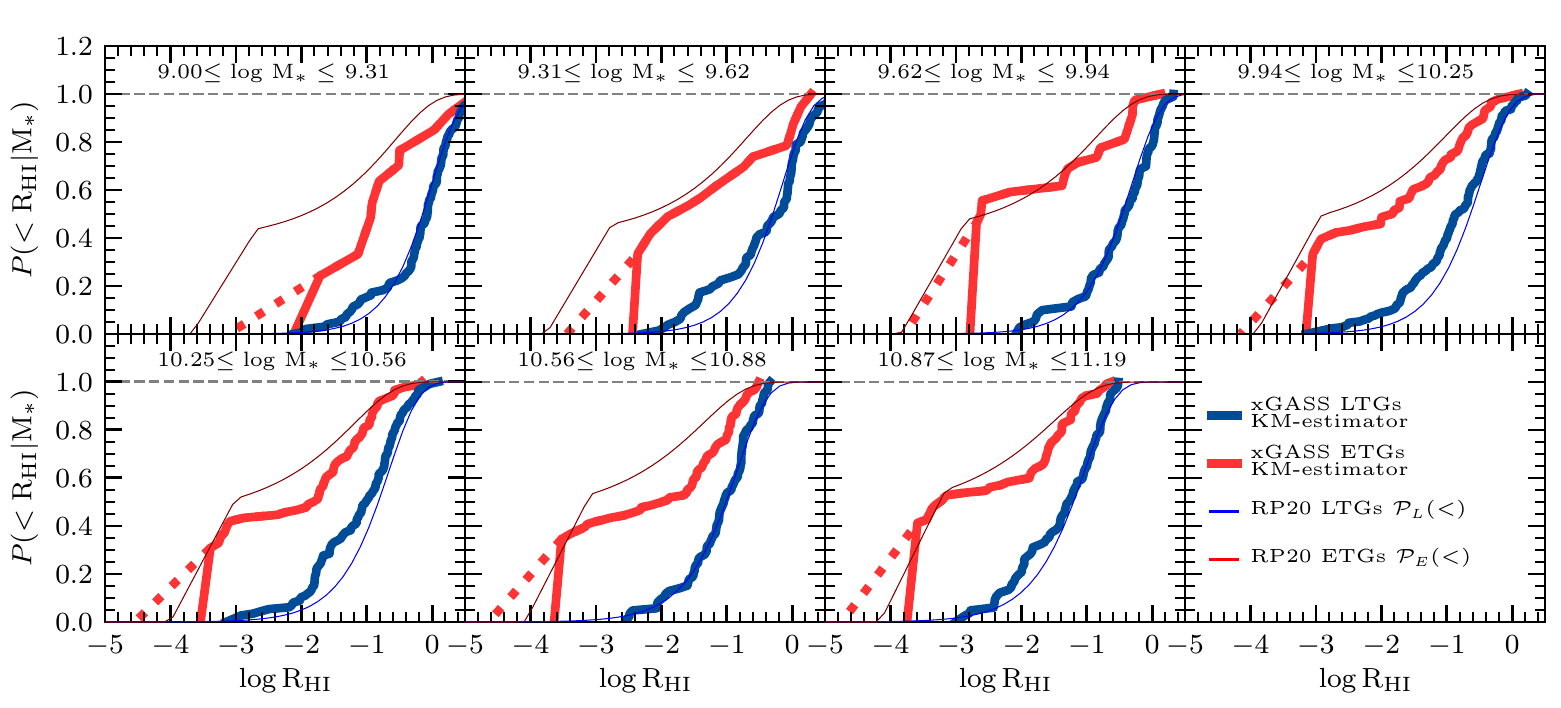} %, height=0.70\textwidth, height=180pt
	\caption{Cumulative histograms of LTG (blue lines) and ETG (red lines) \HI\ conditional distributions (CDFs) at different \ms\ bins from the processed \xG\ sample. For comparison, fits to the respective CDFs from Paper II are shown with thin lines. For a correct comparison, these fits were averaged within the mass ranges of the bins. 
	} 
	\label{fig:PDFs-comparison} 
\end{figure*}
%%%%%%%%%%%%%% END FIGURE%%%%%%%%%%%%%

\subsection{Conditional HI distributions for all, central, and satellite galaxies}

In Figure \ref{fig:PDFs-comparison}, we compare the \RHI\ conditional CDFs of late- and early-type galaxies from the processed \xG\ sample (thick solid lines)  with those inferred empirically in Papers I and II (thin solid lines). The fits were averaged within the width of the \ms\ bin. 
The cumulative distributions for \xG\ ETGs start at fractions around $0.3-0.4$.
These are the fractions of the remaining upper limits after our corrections of \S\S \ref{sec:upper-limits} and \ref{KM-estimator}.
If we proceed as in Paper I, we should assign \RHI\ values following a top-hat function of width $\sim 1$ dex below the lowest upper limit value in each mass bin  for undetected ETGs. This is shown in Figure \ref{fig:PDFs-comparison} with dotted lines.
The \xG\ \HI\ conditional CDFs for LTGs agree well with the analytical fits constrained in Paper I. For ETGs, the CDFs from \xG\ tend to be somewhat shifted to higher \RHI\ values than those determined in Paper I.\footnote{In Paper I, to infer the \HI\ conditional distributions, (i) we used not only the \texttt{GASS} survey but other samples, and (ii) for converting to detections a fraction of the ETG \texttt{GASS} upper limits, a uniform \RHI\ distribution was used while here the empirical \RHI\ conditional PDFs for ETGs constrained in Paper I are used, see Appendix \ref{ETG-corrs}. Therefore, we expect differences between the \HI\ conditional CDFs of ETGs in Paper I and those estimated here for \xG. }
Differences are seen also in the respective logarithmic mean values plotted in Figure \ref{fig:xGASS-correlations}. 

In Appendix \ref{App:no-corrections} we compare the \RHI{} conditional CDFs shown in Fig. \ref{fig:PDFs-comparison} with those obtained without correcting the \xG\ upper limits, Fig. \ref{fig:CDFs-comparison}. From this comparison, it is evident that without this correction, the CDFs for ETGs result poorly constrained. 

\begin{figure*}
	%trim=l b r t
	\includegraphics[trim = 4mm 46.5mm 0mm 5mm, clip, width=0.9\textwidth]{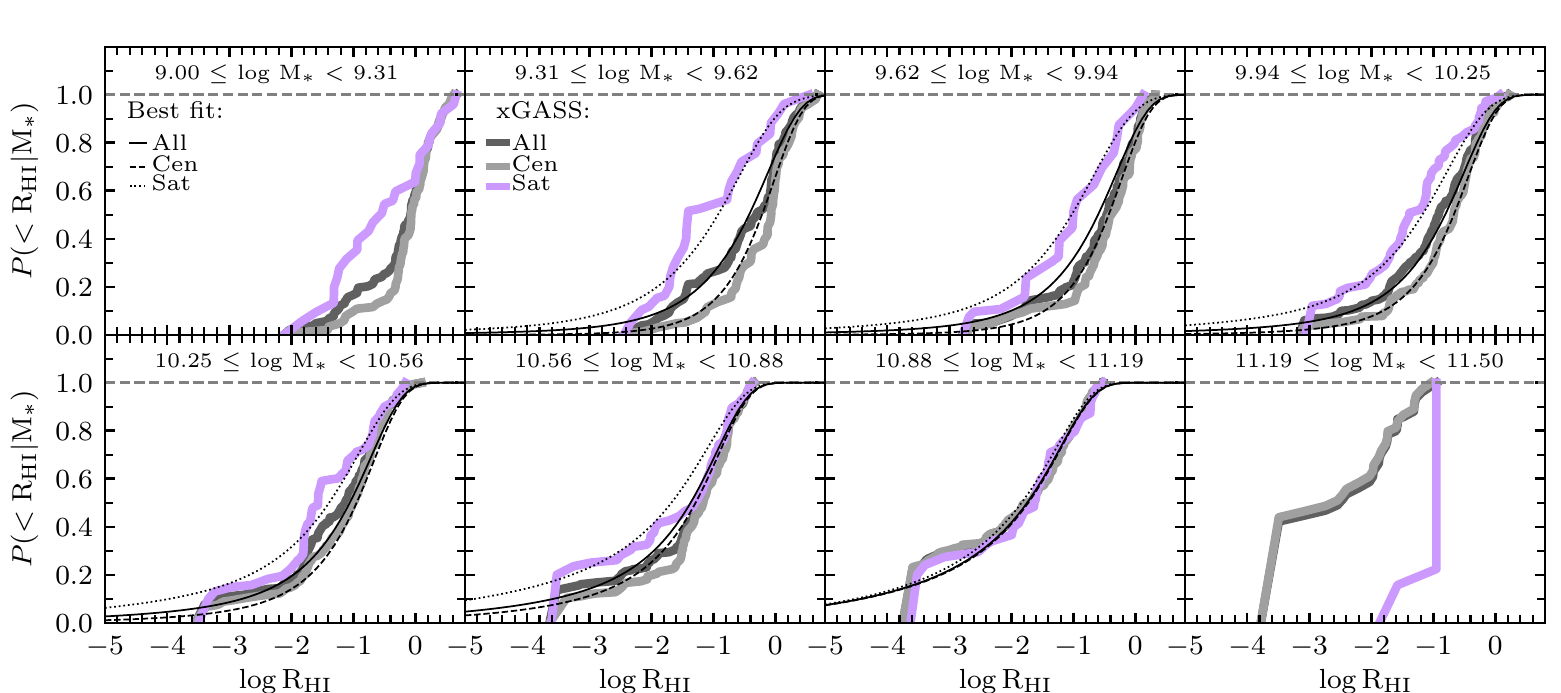} %, height=0.70\textwidth, height=180pt
	\caption{Cumulative \HI\ conditional distributions (CDFs) at different \ms\ bins from the processed \xG\ sample of all galaxies and for only centrals and satellites, see colour notation in the first panel. The solid, dashed, and dotted lines are our best joint fits to the different subpopulations shown in this figure and in Figures \ref{fig:xGASS-CDFs-LTGs} and \ref{fig:xGASS-CDFs-ETGs}, see text.
	} 
	\label{fig:xGASS-CDFs} 
\end{figure*}

\begin{figure*}
	%trim=l b r t
	\includegraphics[trim = 4mm 46.5mm 0mm 5mm, clip, width=0.9\textwidth]{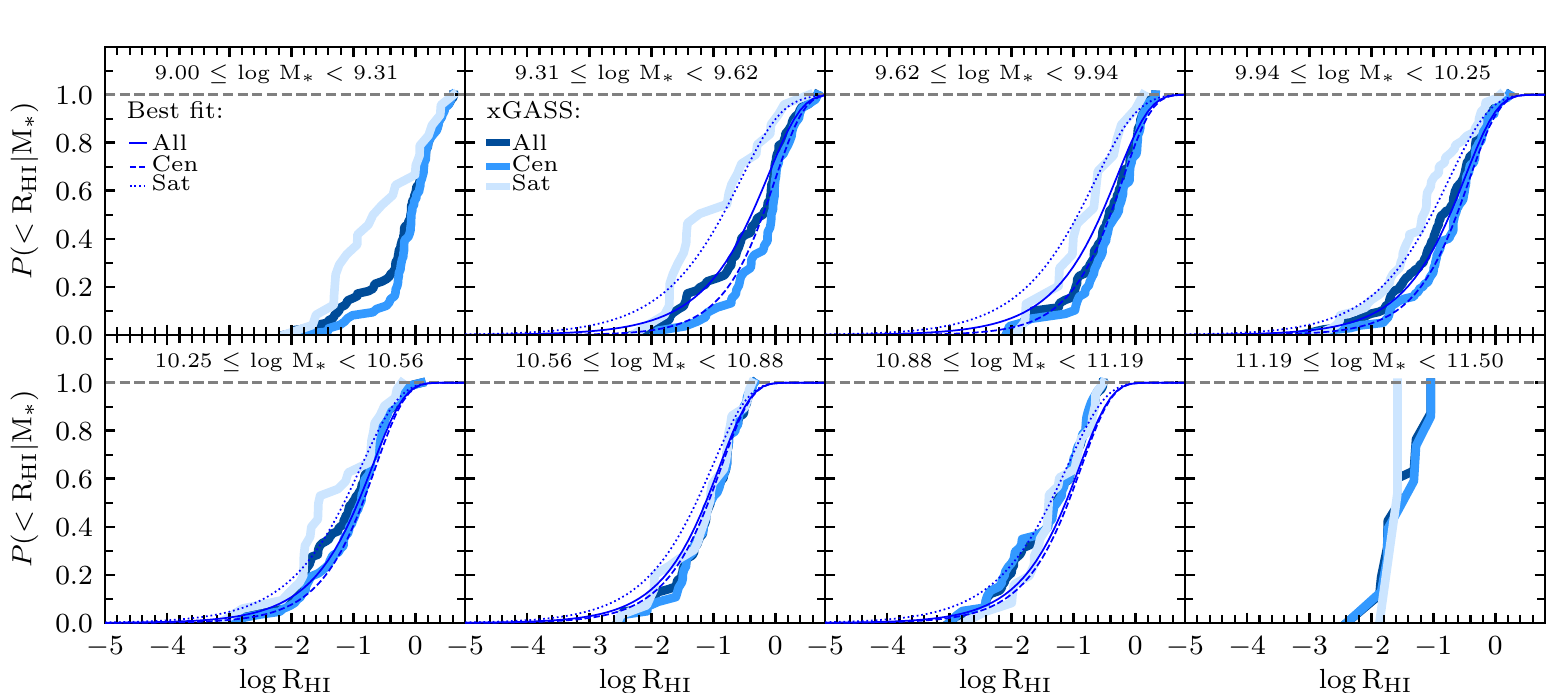} %, height=0.70\textwidth, height=180pt
	\caption{As Figure \ref{fig:xGASS-CDFs} but for the subsample of LTGs.} 
	\label{fig:xGASS-CDFs-LTGs} 
\end{figure*}

\begin{figure*}
	%trim=l b r t
	\includegraphics[trim = 4mm 46.5mm 0mm 5mm, clip, width=0.9\textwidth]{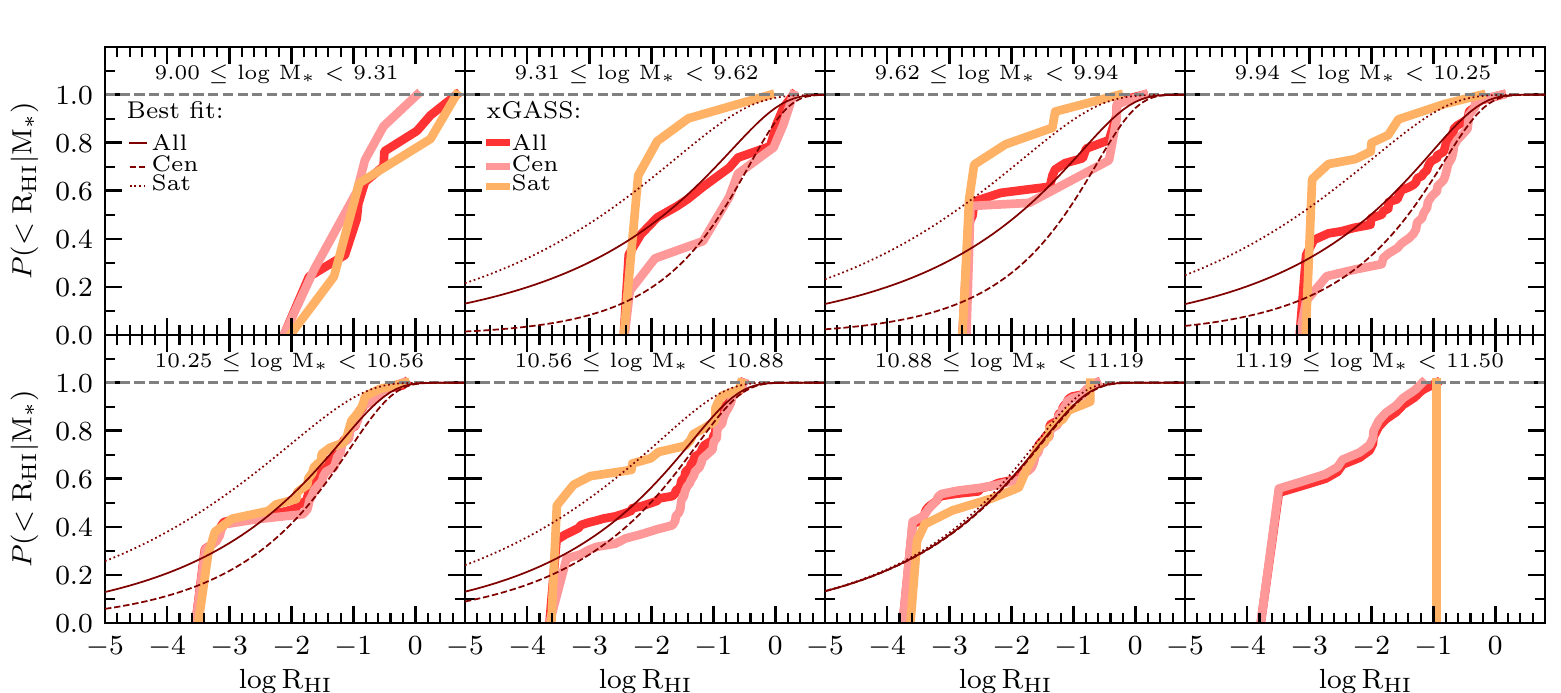} %, height=0.70\textwidth, height=180pt
	\caption{As Figure \ref{fig:xGASS-CDFs} but for the subsample of ETGs.} 
	\label{fig:xGASS-CDFs-ETGs} 
\end{figure*}

Figure \ref{fig:xGASS-CDFs} presents the \RHI\ conditional CDFs in different \ms\ bins calculated as described in \S\S \ref{KM-estimator} for the whole \xG\ sample (black lines), and for centrals (dark grey lines) and satellites (purple lines) only, that is, $P^{i}(>\RHI|\ms)$, where $i$ refers to all, central or satellite, respectively. We find that the lower \ms\ the larger the difference in the distributions between central and satellite galaxies, with the latter having lower \RHI. 
Recall that for calculating these distributions, the \xG\ sample has been weighted by morphology and environment to agree with the SDSS DR7 fractions as a function of \ms, see \S\S \ref{subsec:morph} and Appendix \ref{app:fractions}. The main bias of \xG\ galaxies is actually by morphology; the bias by environment is small and mainly due to the former.
Figures \ref{fig:xGASS-CDFs-LTGs} and \ref{fig:xGASS-CDFs-ETGs} are as Figure \ref{fig:xGASS-CDFs} but now for LTGs and ETGs from \xG, respectively.

%%%%%%%%%%%%%%%%FIGURE%%%%%%%%%%%%%%%
%%%%%%%%%%%%%% END FIGURE%%%%%%%%%%%%%

 %%%%%%%%%%%%%%%%FIGURE%%%%%%%%%%%%%%%
%%%%%%%%%%%%%% END FIGURE%%%%%%%%%%%%%

 %%%%%%%%%%%%%%%%FIGURE%%%%%%%%%%%%%%%
%%%%%%%%%%%%%% END FIGURE%%%%%%%%%%%%%

%%%%%%%%%%%%%%%%%%%%%%%%%%%%%%%%%%%%%%%%%%%%%%%%%%%%%%%%%%
%TABLE: 
\begin{table}
	\centering
	\caption{Best-fitting parameters to four sets of \HI\ CDFs} 
	\resizebox{8.5cm}{!} {
		\begin{tabular}{cccccc}
			\hline
			\hline
			CDFs  & $a$   &  $b$  &  $c$  & $e$  \\ \hline
			LTGs & {0.005} $\pm$ {0.09} &  {0.53} $\pm$ {0.09} & {0.79} $\pm$ {0.18} & {0.67} $\pm$ {0.12} \\ 
			LTGs Centrals & {-0.21} $\pm$ {0.15} &  {0.71} $\pm$ {0.15} & {0.67} $\pm$ {0.12} &  {0.60} $\pm$ {0.11}\\
			ETGs & {0.07} $\pm$ {0.05} &  {0.22} $\pm$ {0.07} & {0.86} $\pm$ {0.09} &  {0.65} $\pm$ {0.09}\\
			ETGs Centrals & {-0.004} $\pm$ {0.11} &  {0.31} $\pm$ {0.15} & {1.09} $\pm$ {0.13} &  {0.75} $\pm$ {0.12}\\ \hline			 
\end{tabular}
	}
	\label{params-CDFs}
\end{table}

%%%%%%%%%%%%%%%%%%%%%%%%END TABLE%%%%%%%%%%%%%%%%%%%%%%%%%

\subsection{Corrections from \xG\ to calculate HI distributions for centrals and satellites}
\label{sec:corrections}

We would like to obtain from the \xG\ analysis presented above a way to estimate the \HI\ conditional CDFs of central and satellite galaxies when only the average CDFs (among centrals and satellites) are known. 
If $\mathcal{P}_j(>\RHI|\ms)$, $j=$LTG or ETG, are the \HI\ conditional CDFs from Paper II, then the corresponding CDFs for central and satellites can be calculated as:
\begin{equation}
    \mathcal{P}_{j}^i(>\RHI|\ms)= \left[\frac{{P}_j^i(>\RHI|\ms)} {{P}_j(>\RHI|\ms)}\right]_{\rm xGASS} \times \mathcal{P}_j(>\RHI|\ms),
   \label{eq:corrections}
\end{equation}
where $i$ refers to either central or satellite galaxy, and the sub-index \texttt{xGASS} refers to analytic fits to the \HI\ CDFs constrained above. Thus, our goal now is to (i) perform a continuous analytic fit to the different \texttt{xGASS} \HI\ CDFs given \ms\ entering in Eq. (\ref{eq:corrections}), and (ii) to be able to extrapolate the fits to lower stellar masses than those of the \texttt{xGASS} sample.

The \HI\ conditional CDFs from the processed \xG\ data presented in Figures \ref{fig:xGASS-CDFs}--\ref{fig:xGASS-CDFs-ETGs} are for the whole sample as well as for different subsamples. In many cases, the numbers of objects in a given \ms\ bin, specially for subsamples containing ETGs and satellites, are low. Then, the CDFs are poorly defined and may suffer of strong sample variance. In view of this, performing fits independently to each CDF is not viable. Besides, it is important that the fitted functions describing the CDFs obey by construction the law of total probability. 
According to this law applied to our context, the relation of the total conditional probability distribution of \RHI\ given \ms, $P_T (<\RHI|\ms)$,  with, for example, two subsamples A and B, with their respective conditional probability distributions $P_A(<\RHI|\ms)$ and  $P_B(<\RHI|\ms)$, is given by: 
\begin{equation}
    P_T (<\RHI|\ms) = P_A(\ms)P_A(<\RHI|\ms) + P_B(\ms)P_B (<\RHI|\ms),
\end{equation}
where $P_A(\ms)$ and $P_B(\ms)$ are the marginalized probability distributions of these subsamples.
In our case, the marginalized probabilities are the fractions of galaxies in the samples $A$ and $B$ as a function of stellar mass, $\phi_A(\ms)/\phi_T(\ms)$ and $\phi_B(\ms)/\phi_T(\ms)$, respectively. 
In Appendix \ref{app:eqs-consv} we present the different equations that should be obeyed according to the law of total probability for the whole sample of galaxies and different subsamples of LTGs/ETGs, centrals/satellites, and their combinations. In these ``probability conservation'' equations enter different fractions of subsamples (the marginalized probability distributions) as a function of \ms. In Appendix \ref{app:fractions} we obtain analytic fits to these fractions using the volume-complete SDSS survey.
As discussed in \S\S\ \ref{subsec:morph}, the fractions of ETGs (centrals or satellites) as a function of \ms\ in \xG\ are different to those from SDSS. This is why we decided to weight the \xG\ sample to agree  with the SDSS DR7 morphological fractions. Having done this, we can use then the SDSS fractions in the ``probability conservation'' equations mentioned above.

Based on the considerations discussed above, we implement the following strategy for obtaining the fits to the \RHI\ conditional CDFs of the whole \xG\ sample as well as of different subsamples:

\begin{enumerate}
    \item Propose parametric functions that describe the \RHI\ conditional CDFs given \ms\ of the following four galaxy subsamples: all LTGs, all ETGs, central LTGs, and central ETGs.  
    \item Calculate the \RHI\ conditional CDFs given \ms\ for: the whole sample of galaxies, and the four subsamples of centrals, satellites, satellite LTGs,  and satellite ETGs, from the CDFs of the previous item by means of the equations of total probability (see Appendix \ref{app:eqs-consv}). 
    \item Implement a continuous joint fitting procedure to the \RHI\ conditional CDFs given \ms\ of the whole sample and the different subsamples mentioned above as obtained from \xG\ after our processing (Figs. \ref{fig:xGASS-CDFs}--\ref{fig:xGASS-CDFs-ETGs}) in order to constrain the parameters of the functions mentioned in the first item. 
\end{enumerate}

For item (i), we propose a generic function for the four subsets of \HI\ CDFs, the incomplete gamma function\footnote{We have shown in Paper I that the \HI\ conditional PDFs given \ms\ can be described by Schechter-like functions. Thus, it is reasonable to propose the incomplete gamma function for describing the respective cumulative PDFs. On the other hand, given the low numbers and non-regular variations in the \RHI\ CDFs with mass of some subsamples from \xG, it is impractical to search for functions with more parameters.}:
 \begin{equation}\label{eq:gamma-tot}
 \mathcal{P}(<x|\ms)=\frac{1}{\Gamma(\alpha)}\int_{0}e^{-x}x^{\alpha-1}dx, 
\end{equation}
where $\Gamma$ is the gamma function,  $x\equiv\RHI/{\rm R_{0}}$, and the parameters $\alpha$ and ${\rm R_{0}}$ depend on $\ms$. We parametrize these dependencies as:
\begin{equation}\label{eq:a-ms}
\alpha(\ms)=a  (\log\ms-10)+b,
\end{equation}
where $a$ and $b$ are the slope and normalization of the power law, respectively, and
\begin{equation}\label{eq:R0-ms}
{\rm R_{0}}(\ms)=\frac{c}{\left(\frac{\ms}{\mtr}\right)^{d}+\left(\frac{\ms}{\mtr}\right)^{e}}.
\end{equation}
Here $c$ is a normalization coefficient, $\mtr$ is the transition mass where the double power law changes its slope, $d$ and $e$ are the slopes for the low- and high-mass ends, respectively. In fact, for the mass range of \xG\ galaxies, a single power law is enough to describe $R_{0}(\ms)$. However, since we will extrapolate the fits of \xG\ \RHI\ CDFs to lower stellar masses, the second power law is necessary. We have found that the values of $d$ and $M_{\rm tr}$ can be fixed, and  not left as free parameters. 
These values were constrained in Paper II from the \HI\ CDFs of LTGs and ETGs for the compilation and processing presented in Paper I in a large \ms\ range; we fix these parameters to the values constrained therein: $d=-0.018$ and $\log(M_{\rm tr}/\msun)=8.646$ for LTGs; $d=-0.820$ and $\log(M_{\rm tr}/\msun)=8.354$ for ETGs. Thus, in Eq. (\ref{eq:gamma-tot}--\ref{eq:R0-ms}) there are four free parameters, $a$, $b$, $c$, and $e$ that remain. The above function Eq. (\ref{eq:gamma-tot}) is proposed to describe each one of the four subsamples of CDFs mentioned in (i). Therefore, we have 16 free parameters in all.

We constrain the 16 free parameters by jointly fitting the nine sets of \RHI\ conditional CDFs from \xG\ mentioned in (i) and (ii) above, and plot them in Figures \ref{fig:xGASS-CDFs}--\ref{fig:xGASS-CDFs-ETGs}. To do so we use a Monte Carlo Markov Chain method described in detail in \citet{Rodriguez-Puebla+2013}. We did not use the information from the largest and lowest stellar mass bins in all the cases because the data in these bins are scarce and the corresponding CDFs are poorly determined. In Table \ref{params-CDFs} we present the best constrained values for the 16 free parameters.  With these values, the four  \xG\ \RHI\ conditional CDFs mentioned in  (i) above are fully described. By using  the equations from Appendix \ref{app:eqs-consv}, the other five \RHI\ CDFs mentioned in (ii) are also described. Thus, any \xG\ \HI\ conditional CDF given \ms\ is described analytically by the fits, in particular those  CDFs in the brackets in Eq. (\ref{eq:corrections}). 
However, we remark that our aim here is not to determine the \RHI\ conditional distributions for the \xG\ survey but to capture the trends with stellar mass of the central- and satellite-to-total ratios as a function of \RHI\ for LTGs and ETGs, that is, the term in the brackets of Eq. (\ref{eq:corrections}). This term combined with our previous accurate inferences of the \RHI\ conditional distributions of LTGs and ETGs (the second term in Eq. \ref{eq:corrections}) will allow us to estimate the respective \RHI\ distributions of central and satellite galaxies. 

The obtained best fits from the continuous joint fitting procedure are shown in Figures \ref{fig:xGASS-CDFs}--\ref{fig:xGASS-CDFs-ETGs} with thin solid, dashed, and dotted lines.  The fits capture the main systematic trends of the different conditional CDFs with \RHI\ and \ms. 
 For some mass bins of ETGs (Fig. \ref{fig:xGASS-CDFs-ETGs}), the fits depart from the data. However, note that the differences between central and satellite galaxy CDFs in these cases move away from the observed overall systematic trend with mass. Recall that the fits are designed to capture the continuous trends for all, late-, and early-type samples {\it jointly}.   While we might propose functions with more parameters, the uncertainties and scarcity of the data for describing the CDFs as a function of \ms\ of the whole sample as well as of the different subsamples do not warranty statistically significant improvements in the fits. 

\begin{figure}
	%trim=l b r t
	%\includegraphics[trim = 6mm 35mm 90mm 13mm, clip, width=0.7\textwidth]{xGASS_logRHI_xGASS.pdf}
	\includegraphics[trim = 10mm 85mm 115mm 8mm, clip, width=\columnwidth]{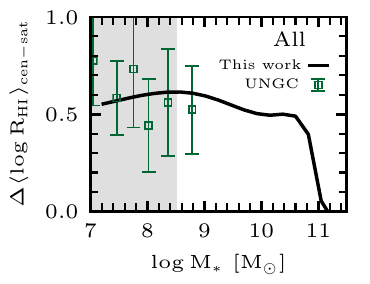} %, height=0.70\textwidth, height=180pt 240pt height=300pt xGASS_logRHI_xGASS.pdf
	\caption{Difference of the logarithmic mean \RHI\ between central and satellite galaxies (in dex). The black solid line corresponds to this difference as a function of \ms\ from our results. Green squares with error bars are differences from the UNGC catalogue for $\ms < 10^{9}$ \msun. Error bars result from propagating the errors of the mean of central and satellites in the given mass bins. The shaded gray area indicates the extrapolation to lower masses of our empirically constrained model.  
	} 
	\label{fig:logRHI-diff} 
\end{figure}

Finally, note that the stellar mass range over which our best-fitting models are constrained for central and satellite galaxies by the \xG\ data is at $10^{9}\lesssim\ms/\msun\lesssim10^{11.5}$. Conservatively, in the next sections, we will assume that our best-fitting models are still valid no more than 0.5 dex above and below the above \ms\ range of the \xG\ data, as indicated in the figures. Nonetheless, our previous empirical determinations for the \RHI\ conditional CDFs (not including the separation between centrals and satellites) extend down to $\ms\sim 10^7$ \msun. 
Thus, to estimate these CDFs separated into centrals and satellites at low masses using Eq. (\ref{eq:corrections}), we extrapolate the best-fitting models constrained by the \xG\ data.  For this, we extrapolate to low masses the constrained mass-dependent functions given in Eqs. (\ref{eq:a-ms}) and (\ref{eq:R0-ms}) as well as the fractions and subfractions as a function of \ms\ entering in the equations of ``probability conservation'' presented in Appendix \ref{app:eqs-consv}. We use the fits to these fractions and subfractions to the SDSS data presented in Appendix \ref{app:fractions} to extrapolate them down to $\ms\sim 10^7$ \msun.  
Unfortunately, information on the \HI\ gas content of dwarf galaxies that have been separated into centrals and satellites is very limited. Such information can be found in the UNGC catalogue of very local galaxies \citep[][]{Karachentsev+2013}, used in Paper I. Figure \ref{fig:logRHI-diff} shows the differences of the logarithmic mean \RHI\ values between centrals and satellites from UNGC (calculated taking into account upper limits) along with these differences as calculated from our \RHI\ conditional distributions and the extrapolations of our best-fitting models to the \xG\ data. The comparison shows that our extrapolation provides results that are consistent within the uncertainties with the UNGC observational data.

%%%%%%%%%%%%%%%%FIGURE%%%%%%%%%%%%%%%

%%%%%%%%%%%%%% END FIGURE%%%%%%%%%%%%%

%========================================
\section{The bivariate \mha\ and \ms\ distributions of central and satellite galaxies}
\label{Sec:results2}
%========================================

%%%%%%%%%%%%%%%FIGURE%%%%%%%%%%%%%%%
\begin{figure*}
%trim=l b r t
\centering
\includegraphics[trim = 10mm 45mm 62mm 8mm, clip, width=\textwidth]{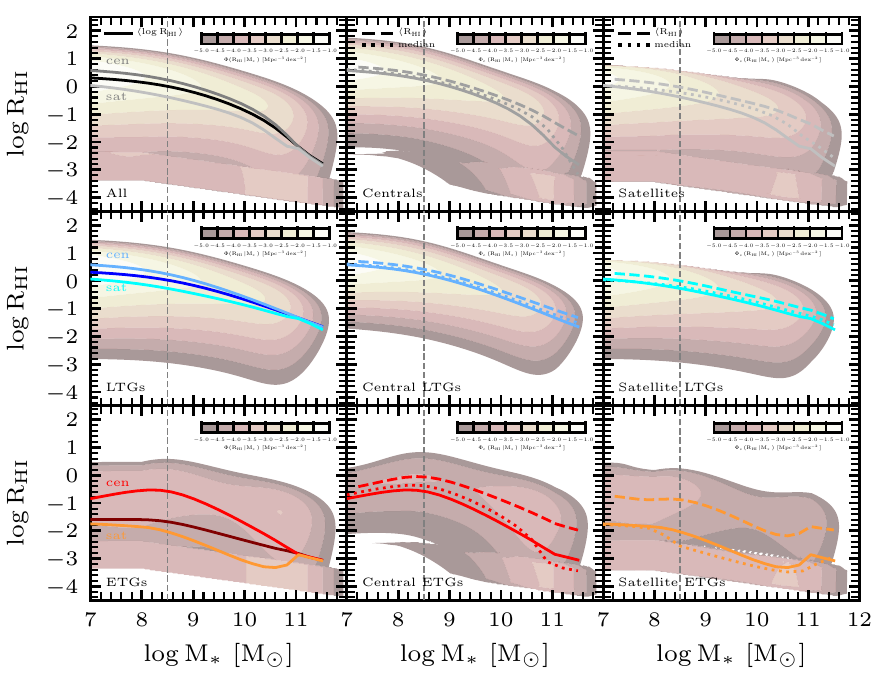} %
\caption{Empirical bivariate \RHI\ and \ms distributions, $\Phi(\RHI|\ms)$. \textit{Upper panels:} From left to right, the distributions for all, central, and satellite galaxies. The solid black, dark gray and light gray lines in the left-hand panel compare the logarithmic means, $\langle\log\RHI\rangle(\ms)$, of all, central, and satellite galaxies, respectively. Satellite galaxies have lower \HI\ gas contents than centrals. The dark and bright lines are reproduced in the medium and right-hand panels, respectively. In these two last panels are also shown the arithmetic means ($\langle\RHI\rangle(\ms)$, dashed line) and the medians (dotted line). \textit{Middle panels:} as the upper panels but now for only LTGs. 
\textit{Lower panels:} as the upper panels but now for only ETGs. The distribution for ETGs is highly bimodal. Hence, the different statistical estimators differ significantly among them. The dashed gray lines indicate extrapolations to lower stellar masses of our empirically constrained model for centrals and satellites.
}
\label{fig:Cen-sat-RHI_Ms} 
\end{figure*}
%%%%%%%%%%%%% END FIGURE%%%%%%%%%%%%%

%%%%%%%%%%%%%%%FIGURE%%%%%%%%%%%%%%%
\begin{figure*}
%trim=l b r t
\centering
\includegraphics[trim = 7mm 70mm 63mm 10mm, clip,  
width=0.95\textwidth]{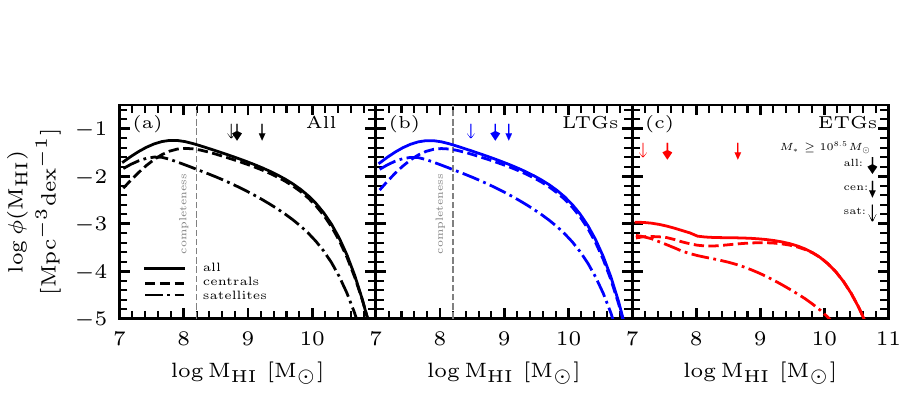} %%,height=160pt ,height=167pt
\caption{\textit{Panel (a):} \HI MF for all, central and satellite galaxies when integrating the bivariate \ms\ and \mha\ distributions over \ms. The shaded green area represents extrapolations for the \HI MF of all galaxies. \textit{Panel (b):} As panel (a) but only for LTGs. \textit{Panel (c):} As panel (a) but only for ETGs. Downward arrows indicate the \HI\ masses corresponding to $\ms\sim 10^{8.5}$, the mass below which our model for centrals and satellites are extrapolations. The vertical dashed lines indicate the completeness limits of our \HI MFs. Due to the low \HI-to-stellar mass ratio of ETGs, note that the \HI\ completeness limit is below $10^7$ \msun.} 
\label{fig:HI-mass-function} 
\end{figure*}
%%%%%%%%%%%%% END FIGURE%%%%%%%%%%%%%

We are now in position to apply the \xG-based functions found in the previous section (and their extrapolations to lower masses) to the \RHI\ conditional distributions of LTGs and ETGs from Paper II to obtain the corresponding distributions for central and satellite galaxies, see Eq. (\ref{eq:corrections}).
The above is the main goal of this paper.
From these \RHI\ distributions as a function of \ms\ we can calculate any statistical estimator, for example the first and second moments, that is, the \RHI--\ms\ relations and their scatters for both central and satellite galaxies. 
In Figures S1-S3 from the supplementary material, we show our empirically determined \RHI\ conditional PDFs for different masses, and for all, late-type, and early-type galaxies separated into centrals and satellites, including our extrapolations to low stellar masses.
In these figures the \xG\ PDFs as obtained in the previous section are also shown. Note than in the latter case, they correspond actually to averages within the given mass bins. 

Following, we extend the results showed in Paper II regarding the joint or bivariate \ms\ and \RHI\ distribution for all galaxies but now separating them into centrals and satellites. As discussed in that paper, by combining the \mha\ (or \RHI) conditional PDFs given \ms\ and the GSMF, $\phi_*(\ms)$, the bivariate distribution function, $\Phi(\RHI,\ms)$, can be calculated. This function is defined as the bivariate number of galaxies within the mass ranges $\log\ms\pm d \log\ms/2$ and $\log\RHI\pm d \log\RHI/2$ in a given volume $V$, and it has units of dex$^{-2}$ Mpc$^{-3}$.

In the left-hand panels of Figure \ref{fig:Cen-sat-RHI_Ms}, from top to bottom, we show the bivariate \ms\ and \RHI\ distribution for all, late-type, and early-type galaxies, respectively. The coloured isocountours correspond to different intervals of bivariate number densities, $\Phi(\RHI,\ms)$, as indicated in the palette (notice that they display four orders of magnitude). 
To construct these bivariate, distributions we used the \RHI\ conditional PDFs given \ms\ for LTGs and ETGs, the GSMF and the fractions of LTGs and ETGs as a function of \ms\ reported in Paper II.
The solid lines show the logarithmic mean relations, $\langle\log\RHI\rangle$-$\log\ms$. 
As extensively discussed in Papers I and II, since LTGs dominate in number density at low masses, the $\langle\log\RHI\rangle$-$\log\ms$ relation of all galaxies is similar to the one of LTGs up to $\ms\sim 10^{10}$ \msun. At higher masses, the fraction of ETGs, which have much lower \HI\ gas contents (compare the medium and bottom left-hand panels of Figure \ref{fig:Cen-sat-RHI_Ms}), increase and then the relation of all galaxies strongly falls to be finally similar to the one of ETGs at $\ms\ga 10^{11.7}$ \msun. 
Note that the \RHI\ distribution for ETGs is non-regular, with a second concentration of galaxies at very low values of \RHI. The above is due to the top-hat component of the \RHI\ conditional PDFs (see Figure \ref{fig:logRHI-logMs-PDFs}). 

The new results from this paper are the bivariate distributions for the galaxies separated into {\it centrals and satellites}, both for the LTG and ETG subsamples as well as for the total galaxy population. The left-hand panels of Figure \ref{fig:Cen-sat-RHI_Ms} also show the logarithmic mean relations for the central and satellite subsamples, respectively.  
The middle and right-hand panels present the bivariate \ms\ and \RHI\ distribution of the central and satellite subsamples with their respective logarithmic mean relations.  
The dashed and dotted lines in these panels show the arithmetic mean relations, $\langle\RHI\rangle$-\ms, and the relations using the median of \RHI, respectively. 

For LTGs, satellites have on average a lower \HI\ gas content than centrals. In particular, 
\HI\ gas-rich galaxies with $\RHI>5$ are all centrals (there are no gas-rich satellites). On the other hand, the gas-poor low-mass LTGs are mostly satellites. At  $\ms\gtrsim 5\times10^{10}$ \msun, central and satellite LTGs have approximately similar \RHI\ gas distributions. 

For ETGs, the difference in the \RHI\ distribution between centrals and satellites is more significant than for LTGs. At $\ms<10^{9}$ \msun, among the ETGs, satellites are much more common than centrals. The \HI\ gas contents of these satellite ETGs is strongly bimodal, with a subpopulation of galaxies with \RHI\ values close to those of the central ETGs and another subpopulation with very low \RHI\ values. For central ETGs of masses $\lesssim 10^{10}$ \msun, there is a small fraction with relatively high values of \RHI. 
They probably correspond to the so-called blue ETGs, some of which are even star forming \citep{Lacerna+2016,Lacerna+2020}. The blue/star-forming ETGs are typically very isolated galaxies and they indeed are expected to have relatively high gas fractions. 
At $\ms>5\times 10^{9}$ \msun, centrals are more common than satellites. The difference in the \RHI\ distribution of the  centrals and satellites ETGs is small.

\subsection{The HI mass functions}

As shown in Paper II, the integration (marginalization) of the bivariate \mha\ and \ms\ distribution over \ms\ results in the \HI MF. The  panel (a)  of Figure \ref{fig:HI-mass-function} presents the above distribution, $\Phi(\mha|\ms)$, for all galaxies and the projected \HI MF (right rotated subpanel). We also plot the logarithmic mean of \mha\ as a function of \ms\ for all, central, and satellite galaxies, as well as the decomposition of the \HI MF into centrals and satellites. For completeness, the GSMFs of all, central, and satellite galaxies are plotted in the upper sub-panel; these functions are actually input in our approach along with the \HI\ conditional PDFs given \ms.

In Paper II it was shown that our empirical \HI MF agrees well with those measured from the blind radio surveys ALFALFA and HIPASS, down to the completeness of our inference, $\mha\sim 10^8$ \msun, which results from the completeness limit of the input GSMF, $\ms \sim 10^7$ \msun. As seen in Fig. \ref{fig:HI-mass-function}, the \HI MF is dominated by central galaxies at all masses. The fraction of centrals (satellites) is $\sim 90\%$ ($\sim 10\%$) or more (less) for $\mha\gtrsim 10^9$\msun. For masses down to $\sim 10^8$ \msun, the fraction of centrals (satellites) decreases down to $\sim 70\%$ (increases up to $\sim 30\%$). The differences in number density between central and satellites are larger for \mha{} than for \ms.  In panels (b) and (c)  of Fig. \ref{fig:HI-mass-function} we present the bivariate distributions and their projections, the \HI MF and GSMF, as in the panel (a), but for the subsamples of LTGs and ETGs. Since LTGs dominante in abundance, their mass functions are similar to those of the whole galaxy population.

\section{Discussion}

\subsection{On the HI gas fraction of central and satellite galaxies}
\label{sec:HI-differences}

There are several pieces of evidence that the \HI\ gas fraction of galaxies tends to be lower in higher-density environments \citep[e.g.,][]{Haynes+1984, Gavazzi+2005,Cortese+2011,Catinella+2013,Rasmussen+2012,Boselli+2014b}.
Studies of the \HI\ gas content of member galaxies within clusters have shown that galaxies in most massive clusters are \HI\ deficient, especially toward the center \citep[e.g.,][]{Haynes+1984,Bravo-Alfaro+2000,Solanes+2001,Serra+2012,Rasmussen+2012,Taylor+2012,Gavazzi+2013}. 
However, the above can be in part due to the morphology-density relation; that is, ETGs, which have exhausted their gas efficiently and early and are {\it intrinsically} gas-poorer, are more abundant in the higher density regions of groups and clusters than LTGs. On the other hand, the \HI\ gas content in very isolated LTGs is on average higher than in cluster LTGs, however, the differences tend to be within the 1$\sigma$ scatter, see Paper I and references therein. The differences between these two opposite environments are larger for ETGs. 

Other authors, rather than exploring environmental effects in specific clusters or for very isolated galaxies, used statistical samples to study the effects of the cluster/group mass and richness on the \HI\ gas content of galaxies, mainly the satellite ones \citep[e.g,][]{Hess-Wilcots2013,Yoon-Rosenberg2015,Stark+2016,Brown+2017,Lu+2020}. 
Once a galaxy becomes a satellite inside a halo, the local environmental effects (ram pressure and viscous stripping, starvation, harassment, tidal interactions) work in the direction of lowering the gas content of the galaxy, more efficiently in more massive and rich halos \citep[see e.g.,][and references therein]{Stark+2016,Stevens+2019}.  It is worth mentioning that in simulations \citep[][]{Wright+2019} it was found that what matters most for the quenching time-scale of satellites is not the halo mass, but the ratio between the satellite galaxy mass to the halo mass, with smaller ratios being associated to faster quenching.

By means of the \HI\ statistical stacking technique applied to an overlap between the ALFALFA survey and the SDSS \citet[][]{Yang+2007} halo-based group catalogue, \citet{Brown+2017} found that satellites in more massive halos have on average lower \HI\ content at fixed \ms\ and specific SFR than those hosted by halos of lower mass. According to their analysis, the systematic environmental suppression of \HI\ at both fixed \ms\ and fixed specific SFR in satellites begins in halo masses typical of the massive group regime ($>10^{13}$ \msun), and fast-acting mechanisms such as ram-pressure stripping are suggested to explain their results. 
\citet[][]{Stark+2016} use RESOLVE, a volume-limited multiwavelength census of $\sim 1500$ local galaxies, to study the \HI-to-stellar mass ratio, \RHI, of satellite galaxies as a function of the halo (group) mass. They found that at fixed \ms, satellites have decreasing \RHI\ values with increasing halo mass at $\mh\gtrsim 10^{12}$ \msun. The analogous relationship for centrals is uncertain and due to the poor overlap in stellar masses between centrals and satellites in the selected halo mass bins, it is not clear how different the \RHI\ values of centrals and satellites are at a fixed \ms. Their results for satellites suggest the presence of starvation and/or stripping mechanisms associated with halo gas heating in intermediate-mass groups. 

The question that we address in this section is how different  the \HI\ gas content between centrals and satellites is at a fixed stellar mass separated explicitly into late- and early-type galaxies. 
In Section \ref{results}, we presented the respective results for the \xG\ survey. Upper limits were corrected for the distance bias (Section \ref{sec:upper-limits}) and included into our survival statistical analysis (Section \ref{KM-estimator}). 
The \RHI\ conditional distributions plotted in Figures \ref{fig:xGASS-CDFs}-\ref{fig:xGASS-CDFs-ETGs} show that they are different among central and satellite galaxies at masses lower than $\sim 3\times 10^{10}$ \msun.  Figure \ref{fig:xGASS-correlations} shows the corresponding $\langle\log\RHI\rangle-\log\ms$ relations of centrals and satellites. 
At fixed \ms, satellites have on average lower \HI\ gas content than centrals with the differences increasing as \ms\ decreases. 
For LTGs, these differences at $\ms\sim10^9$ \msun\ are of $\sim 0.6$ dex, decreasing  to 0 at masses $\ms\sim10^{11}$ \msun. For ETGs, the differences are of $\sim 1$ dex at masses $\ms\lesssim 10^{10}$ \msun. However, it should be noted that the scatter (standard deviation) around the mean relations of centrals and satellites is large and the differences between the corresponding relations of both populations is smaller than their standard deviation.

%%%%%%%%%%%%%%%%FIGURE%%%%%%%%%%%%%%%
\begin{figure*}
	%trim=l b r t
	\includegraphics[trim = 5mm 65mm 60mm 8mm, clip, width=0.95\textwidth]{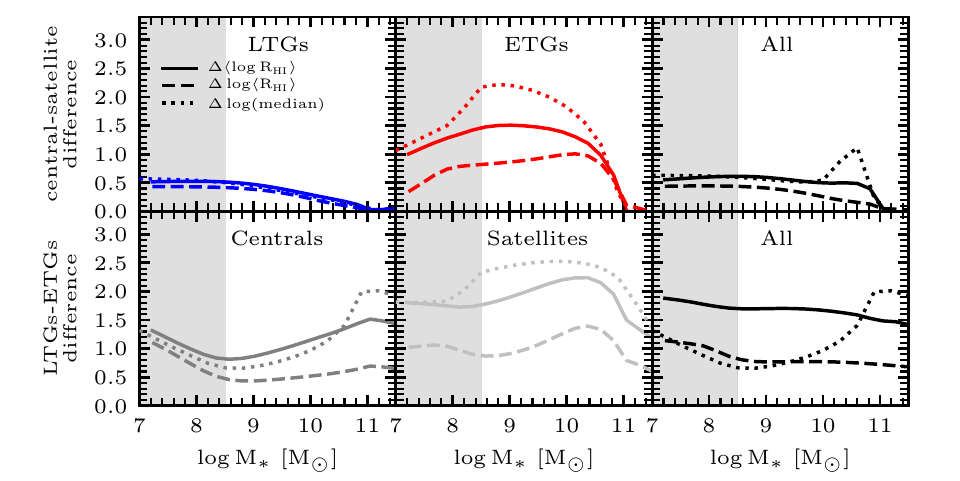} %, height=0.70\textwidth, height=180pt
	\caption{\textit{Upper panels:} Relative differences (in dex) between the logarithmic means (solid line), arithmetic means (dashed line), and medians (dotted line) of centrals and satellites as a function of \ms. From left to right, these relative differences are shown for the LTG, ETG, and whole galaxy populations. In all the cases, centrals have higher \HI\ gas contents than satellites, but at the largest masses, these differences become very small. \textit{Lower panels:} As the upper panels but in this case the relative differences are between LTGs and ETGs for the central, satellite and whole galaxy population (from left to right, respectively). LTGs have much higher \HI\ gas contents than ETGs at all masses. In both, upper and lower panels shaded gray areas indicate extrapolations to lower stellar masses of our empirically constrained model for centrals and satellites.
	} 
	\label{fig:delta-RHI-ave-median} 
\end{figure*}
%%%%%%%%%%%%%% END FIGURE%%%%%%%%%%%%%

By using the \xG\ measurements to the \HI\ conditional distributions,  in Section 3.3 we constrained a set of proposed functions that allow us to project the \RHI\ conditional PDFs for LTGs and ETGs presented in Paper II into their corresponding distributions of centrals and satellites. 
The obtained bivariate (joint) \RHI\ and \ms\ distributions are shown in Figure \ref{fig:Cen-sat-RHI_Ms} along with their respective relations using different statistical estimators.  
As discussed in Section \ref{Sec:results2}, the bivariate distributions of centrals and satellites are different for both LTGs and ETGs, and consequently for all galaxies.  The differences depend on mass and for ETGs they are not easy to quantify by statistical estimators due the non-regular distribution of \RHI.

To dig deeper into the differences in our empirically constructed \HI\ distributions of central and satellite galaxies, we apply a two-sample Kolgomorov-Smirnov test to the obtained \RHI\ conditional PDFs given \ms\ of centrals and satellites for late-type, early-type and all galaxies (Figures S1-S3 in the supplementary material). 
Quantitatively, the central and satellite \HI\ distributions are different at the 95\% or higher level ($p<0.05$) for $\ms\lesssim 3\times 10^{10}$ \msun\ in all the cases.
For larger masses, the differences are smaller and both centrals and satellites are consistent with being drawn from the same distribution of \HI\ gas content.

In the upper panels of Figure \ref{fig:delta-RHI-ave-median}, the relative differences in $\langle\log\RHI\rangle$ (solid lines) and median \RHI\ (dotted lines) between centrals and satellites as a function of mass are shown for late-type, early-type, and all galaxies. We show also the arithmetic mean, $\langle\RHI\rangle$. For LTGs,  the relative difference between centrals and satellites is negligible at masses around $10^{11}$ \msun\ and it increases up to $\sim 0.55$ dex at $\ms\sim 5\times 10^8$ \msun, remaining similar at lower masses. The relative differences for the arithmetic mean are slightly smaller than for the logarithmic mean or the medians. 
For ETGs, the relative differences between centrals and satellites are larger than for LTGs. Since for ETGs, and for both centrals and satellites, the \RHI\ conditional distributions given \ms\ are non-regular, the statistical estimators (geometric or arithmetic mean and median) significantly differ among each other, and consequently, also different is the relative difference among these estimators for centrals and satellites. Our results suggest that the relative difference in $\langle\log\RHI\rangle$ is negligible for $\ms>10^{11}$ \msun, but at lower masses, satellites are much more \HI\ gas-poor than centrals, by $\sim 1.2$ dex at $\ms\sim3\times 10^{8}-5\times 10^{9}$ \msun. 
The relative difference in the medians, is larger than in the logarithmic means, specially at the range $\ms\sim3\times 10^{8}-10^{10}$ \msun.
For the arithmetic means, the relative difference is significantly lower at all masses. The arithmetic means minimize the contribution of galaxies with very low \RHI\ values, which in the case of ETGs, as already discussed, distribute in a dominant second mode both for central and satellite galaxies (see their \RHI\ conditional PDFs in Fig. S3 from the supplementary material). 
For the combined population of late- and early-type galaxies, the relative differences in $\langle\log\RHI\rangle$ between centrals and satellites are $0.4-0.6$ dex for $\ms<5\times 10^{10}$ \msun. The differences are slightly larger for the medians and smaller for the arithmetic means.

Finally, from Figure  \ref{fig:Cen-sat-RHI_Ms} we note that the \HI\ distributions of late- and early-types (left-hand panels) differ much more than the distribution of centrals and satellites (top panels). 
The lower panels of Figure \ref{fig:delta-RHI-ave-median} show the relative differences in $\langle\log\RHI\rangle$ and median \RHI\ between LTGs and ETGs as a function of mass for central, satellite and all galaxies, from left to right, respectively. We also show the respective differences but for the arithmetic mean, $\langle\RHI\rangle$, as dashed line. The relative differences in the lower panels are much higher than in upper panels.
Overall, the above can be interpreted as the present-day \HI\ gas content of galaxies depending more on their internal nature, that is, whether they are of late or early type morphology, than on external conditions associated to whether the galaxy is central or satellite. 
Nevertheless, this claim should be taken with caution. As mentioned above, there is evidence of the \HI\ gas content of satellite galaxies being lower in massive haloes than in less massive ones at fixed stellar mass. 
It is interesting to mention that internal galaxy properties such as colour or specific star-formation rate could correlate even better with the \HI\ gas content than morphology. For instance, \citet[][]{Cook+2019} showed that selecting only the subset of star-forming galaxies in the \xG\ sample, the observed dependence at a fixed \ms\ of \HI\ gas content on bulge-to-total ratio (a proxy for morphology) tends to disappear. 
The dependence of \HI\ gas content on either internal properties, such as morphology, or on external conditions, such as the galaxy being central or satellite, could be related to both if the environment is responsible for reducing the gas content --and consequently quenching the star formation-- and morphologically transforming galaxies. However, while common environmental effects such as ram pressure and starvation drain the gas and quench the satellites, a morphological transformation is not expected \citep[e.g.,][]{vandenBosch+2008,Weinmann+2009}. There is evidence that perhaps low-mass discs can be transformed into S0 gas-poor galaxies when they fall into clusters of galaxies, while the formation of massive S0 galaxies seems to be more related to high-redshift dissipational processes \citep[][and more references therein]{Fraser-McKelvie+2018}.

\subsection{Caveats}

\subsubsection{Effects of different morphological classifications}
\label{sec:DS18}

The results presented here partially depend on the adopted criteria to morphologically classify galaxies as late- or early types. According to the above, we have separated the \xG\ sample into LTGs and ETGs, and estimated the different fractions and subfractions as a function of \ms\ (Appendix \ref{app:fractions}) required for our fitting procedure, by using the automatic morphological classification of \citet{Huertas-Company+2011} implemented for SDSS galaxies. Next, we explore how much our results are affected by using an alternative morphological classification. \citet[][]{Dominguez-Sanchez+2018} applied an automatic classification method to determine the morphology of the SDSS galaxies. We use their results to separate the \xG\ sample into LTGs and ETGs, by employing the same morphological division criterion as we did in the case of the 
\citet{Huertas-Company+2011} classification. Recall that elliptical and S0 galaxies were defined as ETGS, and from Sa to later types as LTGs. 

%%%%%%%%%%%%%%%FIGURE%%%%%%%%%%%%%%%
\begin{figure}
%trim=l b r t
\centering
\includegraphics[trim = 5mm 71mm 57mm 12mm, clip, width=\columnwidth,height=120pt]{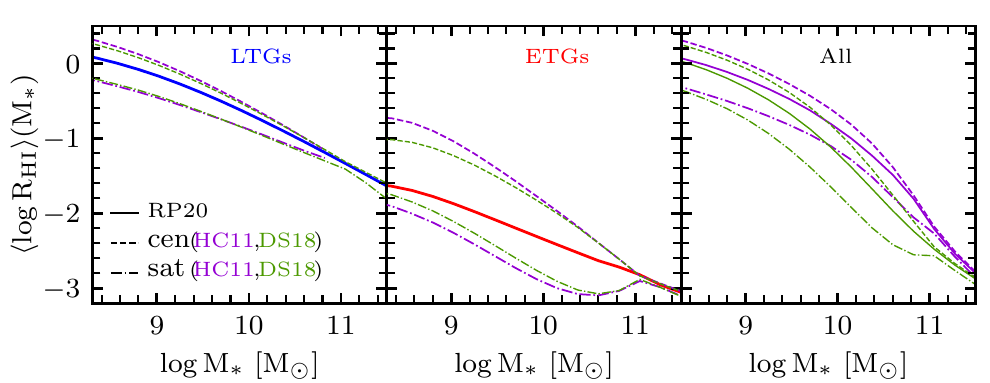} %
\caption{Logarithmic mean \RHI--\ms\ relations for LTGs, ETGs and all galaxies (solid lines) and their respective decomposition into central (short-dashed lines) and satellite (dot-dashed lines) using the  \citet[][magenta; see also Fig. \ref{fig:Cen-sat-RHI_Ms}]{Huertas-Company+2011} and \citet[][green]{Dominguez-Sanchez+2018} morphological classifications.} 
\label{fig:HC-vs-DS} 
\end{figure}
%%%%%%%%%%%%% END FIGURE%%%%%%%%%%%%%

The \citet[][]{Dominguez-Sanchez+2018} morphological classification finds more ETGs than the one from \citet{Huertas-Company+2011} at all masses, see Appendix \ref{app:fractions}. As a consequence, the fractions of the different subpopulations change in \xG, and also change the \HI\ conditional CDFs corresponding to these subpopulations.  We have repeated the whole analysis presented in \S \ref{results} but for the \citet[][]{Dominguez-Sanchez+2018} morphological classification, and obtained different functions for the $[P_i^j(>\RHI|\ms) /{P}_i(>\RHI|\ms)]_{\rm xGASS}$ ratios appearing in Eq. (\ref{eq:corrections}). 
By using these new functions, we calculated the corresponding \HI\ CDFs of central and satellite galaxies for the LTG and ETG populations. 
Notice that the weights applied to \xG\ were changed accordingly.

Figure \ref{fig:HC-vs-DS} compares the resulting mean $\langle\log\RHI\rangle$-$\log\ms$ relations of centrals and satellites for the LTG and ETG populations from the \citet{Huertas-Company+2011} and the \citet{Dominguez-Sanchez+2018} morphological classifications. 
The differences in the $\langle\log\RHI\rangle$-$\log\ms$ relations of centrals and satellites introduced by the uncertainty in morphological classification are negligible for LTGs. These differences for ETGs range from $\sim 0.35$ to 0.05 dex at masses  $\sim 2\times 10^8$ and $\ms\sim 2\times10^{10}$ \msun, respectively with \citet{Dominguez-Sanchez+2018} classification giving less separation into centrals and satellites than the \citet{Huertas-Company+2011} one.  
At higher masses, differences between one or another classification scheme are negligible.
The total relations shown in the right-hand panel are the weighted averages of LTGs and ETGs. Recall that the weights applied to xGASS depend on the morphological classification scheme, see Appendix \ref{app:fractions}. This is why the total $\langle\log\RHI\rangle$-$\log\ms$ relation is different when using one or the other morphological classification. Since for the \citet{Dominguez-Sanchez+2018} classification the fraction of ETGs is larger than for the \citet{Huertas-Company+2011} classification, and because ETGs are \HI\ gas poorer than LTGs, the mean  $\langle\log\RHI\rangle$-$\log\ms$ relation in the former case is below than in the latter case. 

In conclusion, variations in the morphological classification affect weakly our inferences of the difference between the \HI\ gas content of centrals and satellites and only for ETGs. Adopting the \citet{Dominguez-Sanchez+2018} morphological classification instead of \citet{Huertas-Company+2011} leads to
a smaller separation in the mean relations of central and satellite ETGs than adopting the latter.

\subsubsection{Effects of membership and central/satellite designation errors}
\label{sec:explorations}

In this paper, we have used the \xG\ survey for modeling the  \HI\ gas content of central and satellites galaxies. As mentioned in \S\S \ref{subsec:morph}, in \xG\ the central/satellite assignation comes from the SDSS group catalogue of \citet[][]{Yang+2005,Yang+2007}. This group catalogue, as others, may suffer of membership allocation and central/satellite designation errors. 
For example, \citet[][]{Campbell+2015} used a group catalogue constructed based on the \citet[][]{Yang+2005} group finder in a galaxy mock sample and estimated that the fraction of satellites that are truly satellites in the mock (purity) is around 70\%, while for centrals, the purity decreases from $\sim 95\%$ at low group masses, $\sim 10^{12}$ $h^{-1}$\msun, to below $60\%$ at masses $\gtrsim 10^{14}$ $h^{-1}$\msun. 
On the other hand, the fraction of satellites in haloes that are correctly assigned to groups (completeness) is $\sim 80\%$ independent of the halo mass, while for centrals the completeness decreases from $\sim 90\%$ at low halo masses to $\sim 60\%$ at the largest masses. The main source of confusion for centrals at large group masses is the central inversion problem, when the most luminous or massive galaxy is a satellite rather than the true central \citep{vandenBosch+2005,Skibba+2011}.  

Thus, the differences in the \HI\ gas fractions between centrals and satellites inferred with \xG\ (see Section \ref{results}) could be larger. The above also implies  that the differences in the overall \HI\ distributions of central and satellites reported in Section \ref{Sec:results2} could be larger. 
Note, however, that for the \xG\ sample that we use here, \citet{Janowiecki+2017} improved the group membership given by \citet[][]{Yang+2007} by visually inspecting false pairs and galaxy shredding. 

\section{Summary and Conclusions}

We have analysed the multiwavelength \xG\ survey \citep[][]{Catinella+2018}, applying the same procedure as in Paper I to 
(i) re-scale their upper limits on the basis of samples observed in radio at lower distances, and (ii) treat the corrected upper limits with a survival analysis in order to infer full statistical distributions of the \HI\ gas content of galaxies. 

We have found that for LTGs, the \RHI--\ms\ relation and the full \RHI\ conditional distributions as a function of \ms\ from \xG\ agree very well with those empirically determined in Paper I for a larger stellar mass range sample (Figure \ref{fig:xGASS-correlations}). For ETGs, the \RHI\ distributions from \xG\ galaxies imply slightly higher values of \RHI\ than our previous determinations. 
For \xG\ LTGs, centrals are on average more \HI\ gas-rich than satellites of the same stellar mass. These differences are negligible for $\log(\ms/\msun)>10.8$, while at the lowest masses, $9.0<\log(\ms/\msun)\lesssim 9.7$, these  differences are 0.5-0.7 dex, on average.  For ETGs, the differences between centrals and satellites are larger than for LTGs. However, in both cases, the 1-$\sigma$ scatter around the \RHI-\ms\ relations of centrals and satellites is larger than the  difference between their means.

By means of a continuous fitting procedure to the processed \xG\ data, we determined a set of functions that allowed us to project our empirical \HI\ conditional cumulative distributions given \ms\ of both LTGs and ETGs into  central and satellite galaxies. In other words, \xG\ provides the information required to estimate the \HI\ conditional distributions of centrals and satellites from the overall \HI\ conditional distributions for both late- and early-type galaxies. 
We use the above mentioned functions to extrapolate to stellar masses lower than those of the \xG\ survey. By combining the \RHI\ conditional distributions given \ms\ with the corresponding GSMFs, the bivariate \ms\ and \RHI\ distribution functions, $\Phi(\ms,\RHI)$, for late-type, early-type, and all galaxies, separated into centrals and satellites, were calculated (Figure \ref{fig:Cen-sat-RHI_Ms}).  The main results obtained from this exercise are summarised below:

\begin{itemize}
    \item For LTGs, satellites have on average less \HI\ than centrals. Up to $\ms\sim 10^9$ \msun, the relative difference is $\sim 0.5$ dex and all the gas-rich dwarf LTGs are centrals. For higher masses, this relative difference decreases up to $\ms\sim 3\times 10^{10}$ \msun, above which there is no difference between centrals and satellites. 
    Since the bivariate distribution is regular for LTGs, even for centrals and satellites separately, the \RHI--\ms\ relations calculated with different statistical estimators are roughly similar.
    
    \item For ETGs, the bivariate distributions for centrals and satellites differ more than for LTGs, satellites being on average more devoid of \HI\ than centrals up to $\ms\sim 5\times 10^{10}$ \msun. However, the \RHI\ distribution of satellite ETGs  is  strongly bimodal, with a fraction of them having \RHI\ values close to those of central ETGs  and  another  fraction with very low \RHI\ values.  At $\ms\gtrsim 5\times 10^{10}$ \msun, central ETGs are already more abundant than satellite ETGs but both have statistically similar \HI\ gas content. 
    
    \item  Since the bivariate distributions for ETGs, both centrals and satellites, are non-regular, the \RHI--\ms\ relations calculated with different statistical estimators are different. In particular, the relation based on arithmetic means, $\langle\RHI\rangle$, is significantly above the relations based on logarihtmic means or medians. 
    
    \item The projection of the bivariate distribution when integrating it over \ms\ is the \HI MF and agrees well with those measured in blind radio surveys. We show here that the \HI MF is completely dominated by central galaxies at all masses, both for LTGs and ETGs (Fig. \ref{fig:delta-RHI-ave-median}). 
    
\end{itemize}

Overall, our results show that the difference in the bivariate \RHI\ and \ms\ distribution between late- and early-type galaxies is significantly larger than between central and satellite galaxies. This suggests that the \HI\ gas content of galaxies depends more on their internal nature, that is, whether they are of late or early type morphology, than on external conditions associated to whether the galaxy is central or satellite.

In this paper we presented a full statistical description of the \HI\ gas content of local galaxies as a function of their stellar mass and separated into late- and early-type and into central and satellites. These results can be used for comparisons with theoretical predictions of galaxy evolution, and for adding the \HI\ gas component in empirical approaches aimed to model the local galaxy population. In particular, our results can be used to establish the \ms-\mha-\mh\ connection from the outcome of large N-body cosmological simulations, where complete mock galaxy catalogues can be generated. In a forthcoming paper, we will present results of this connection including predictions on the spatial clustering of galaxies using both their stellar and \HI\ masses. 

\section*{Acknowledgements}
The authors thank the anonymous Reviewer for her/his comments and suggestions that helped to improve the presentation of this paper. 
ARC acknowledges CONACyT for a PhD fellowship. ARP and VAR acknowledge financial support from CONACyT through ``Ciencia Basica'' grant 285721, and  from DGAPA-UNAM through PAPIIT grant IA104118.
CL and BC acknowledge partial funding from the ARC Centre of Excellence for All Sky Astrophysics in 3 Dimensions (ASTRO 3D), through project number CE170100013. 

\section*{Data availability}
The xGASS, the automated morphological classification for SDSS DR7 and the \citet[][see also \citealp{Yang+2007}]{Yang+2012} galaxy group catalogues are publicly available\footnote{xGASS: \url{https://xgass.icrar.org/data.html}, \\ Automated morphological classification for SDSS DR7: \url{http://cdsarc.u-strasbg.fr/viz-bin/qcat?J/A+A/525/A157}, \\ and \cite{Yang+2012} galaxy group catalogues: \url{https://gax.sjtu.edu.cn/data/Group.html}  }. The data underlying this article will be shared on reasonable request to the corresponding author.

%%%%%%%%%%%%%%%%%%%%%%%%%%%%%%%%%%%%%%%%%%%%%%%%%%



\bibliographystyle{mnras}
\bibliography{references} % if your bibtex file is called example.bib

%\begin{thebibliography}{99}
%\end{thebibliography}

%%%%%%%%%%%%%%%%%%%%%%%%%%%%%%%%%%%%%%%%%%%%%%%%%%

%%%%%%%%%%%%%%%%% APPENDICES %%%%%%%%%%%%%%%%%%%%%

\appendix

\section{Correcting \xG\ to the morphology and environment distributions of SDSS}
\label{app:fractions}

In this Appendix, we first define the fractions corresponding those galaxy subsamples required to perform the joint analytic fitting to the \xG\ \HI\ conditional CDFs in \S\S \ref{sec:corrections}. Then, we compare these fractions from \xG\ to those from the volume-corrected SDSS DR7. Finally, we explain our procedure for weighting \xG\ galaxies in order they reproduced the fractions of ETGs  %\citep[bo][classification]{Huertas-Company+2011} 
and of satellites %\citep[][central/satellite designation]{Yang+2007} 
from SDSS.
Following, in the definition of the different fractions, for simplicity, we omit the dependence on \ms. %We begin by defining the following fractions:
\begin{itemize}
    \item[(i)] \textit{Fraction of ETGs/LTGs:}\\
    Defined as the ratio of ETG to total mass functions, $f^{E}\equiv\phi^{E}/\phi$. The fraction of LTGs, $f_{L}$, is the complement, $f^{L}=1-f^{E}$.
    
    \item[(ii)] \textit{Fraction of centrals/satellites}:\\
    Defined as the ratio of central to total mass functions, $f^{c}\equiv \phi^{c}/\phi$. The fraction of satellites, $f^{s}$, is the complement, $f^{s}=1-f^{c}$.
    
    \item[(iii)] \textit{Fraction of ETGs/LTGs for centrals}:\\
    For the subsample of centrals described by the central mass function $\phi^{c}$,  $f^{c}_{E}\equiv \phi^{c}_{E}/\phi^{c}$ is the fraction of ETGs for centrals. The respective fraction of LTGs is the complement, $f^{c}_{L}=1-f^{c}_{E}$.
    
    \item[(iv)] \textit{Fraction of ETGs/LTGs for satellites}:\\
    For the subsample of satellites described by the satellite mass function $\phi^{c}$, $f^{s}_{E}\equiv \phi^{s}_{E}/\phi^{c}$ is the fraction of ETGs for satellites. The respective fraction of LTGs is the complement, $f^{s}_{L}=1-f^{s}_{E}$.

    \item[(v)] \textit{Fraction of satellites/centrals for LTGs}:\\
    For the subsample of LTGs described by the LTG mass function $\phi^{L}$, 
    $f_{s}^{L}\equiv \phi_{s}^{L}/\phi^{L}$ is the fraction of satellites for LTGs.
    The respective fraction of centrals is the complement, $f_{c}^{L}=1-f_{s}^{L}$. 
    
    \item[(vi)] \textit{Fraction of satellites/centrals for ETGs}:\\
    For the subsample of ETGs described by the ETG mass function $\phi^{E}$,
    $f_{s}^{E}\equiv \phi_{s}^{E}/\phi^{E}$ is the fraction of satellites for ETGs. The respective fraction of centrals is the complement, $f_{c}^{E}=1-f_{s}^{E}$.
\end{itemize}

%%%%%%%%%%%%%%%%FIGURE%%%%%%%%%%%%%%%
\begin{figure*}
	%trim=l b r t
	\includegraphics[trim = 3mm 43mm 0mm 12mm, clip, width=\textwidth]{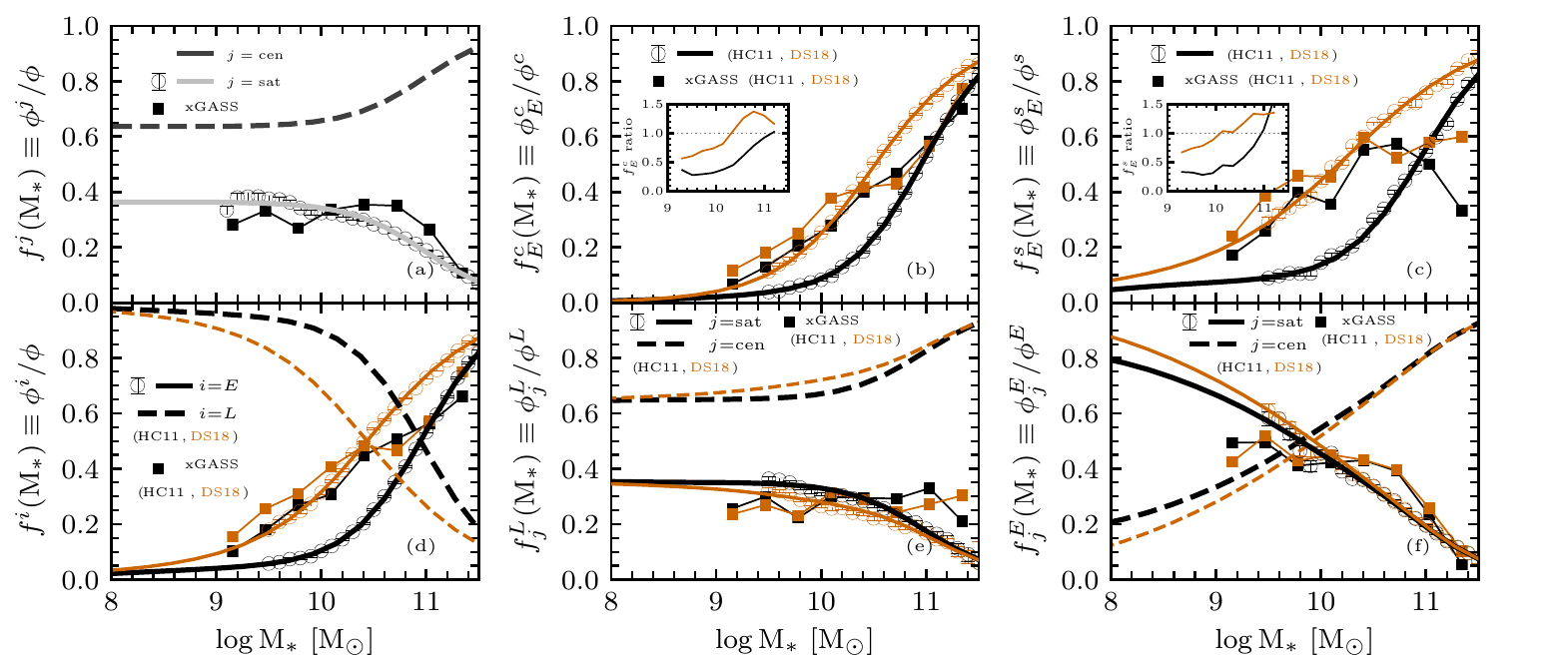} %, height=0.70\textwidth, height=180pt
	\caption{Different fractions of subsamples calculated from the volume-complete SDSS using the \citet[][]{Yang+2012} group catalogue for defining centrals and satellites. Black and brown colours are inferences using the \citet[HC11,][]{Huertas-Company+2011} and \citet[DS18,][]{Dominguez-Sanchez+2018} morphological classifications, respectively. The fractions corresponding to \xG\ are shown with filled squares connected by solid lines. The solid lines are fits to the SDSS data; the dashed lines show the respective complementary fractions. The insets in panels (b) and (c) are the ratios of SDSS to \xG\ fractions.
	}
	\label{fig:fractions} 
\end{figure*}
%%%%%%%%%%%%%% END FIGURE%%%%%%%%%%%%%

Figure \ref{fig:fractions} shows the fractions defined above for \xG, solid black squares connected with solid lines. From left to right, the upper panels show the fraction of satellites for all galaxies (ii), and the fractions of ETGs for the subsamples of central and satellite galaxies (iii and  iv). The lower panels show the fraction of ETGs for all galaxies (i), and the fractions of satellites for the subsamples of LTGs and ETGs (v and  vi). In these panels, the respective fractions measured from the volume-complete SDSS DR7 are also plotted (black circles) along with analytical fits to these fractions (black lines; see below).
We use \cite{Meert+2015} photometry and an average stellar mass from five different mass-to-luminosity prescriptions, updated galaxy group catalogues from \cite{Yang+2007,Yang+2012}, and the \cite{Huertas-Company+2011} morphological classification (see Paper II for details and for the corrections applied to obtain a volume complete sample). As seen in panels (b) and (c), the fractions of ETGs in the central and satellite subsamples are systematically larger up to $\ms\sim 10^{11}$ \msun\ for \xG\ than for SDSS; at larger masses, the difference inverts for the subsample of satellite galaxies. In the insets of these panels, we plot the ratios of the respective fractions of SDSS to \xG. 

As mentioned in \S\S \ref{subsec:morph}, to infer from \xG\ the \RHI--\ms\ relations and \RHI\ distributions given \ms\ corresponding to all galaxies, as well as to the subsamples of central and satellite galaxies, the biases of \xG\ with respect to SDSS in morphology and environment should be corrected. To do so, we adopt a methodology similar as in \citet[][]{Catinella+2018} for recovering a volume complete sample. When we compute the above mentioned \RHI--\ms\ relations or the whole \RHI\ conditional distributions given \ms, we apply weights to \xG\ galaxies to recover the volume-compete SDSS fractions of ETGs in the central and satellite subsamples.  The weights are the ratios shown in the insets of panels (b) and (c) of Figure \ref{fig:fractions}. This automatically also recovers the overall SDSS fraction of ETGs and the overall fraction of satellites. In any case, note that the relevant bias of the \xG\ sample with respect to SDSS is by morphology; the bias in selecting central/satellite galaxies is small and mainly due to the former.

For the above procedure and for extrapolating the fits to the \RHI\ conditional CDFs from \xG\ to masses lower than $\ms=10^9$ \msun, we use actually analytical fits to the SDSS fractions.  The fits are performed to the overall fraction of satellites (panel a) and the fractions of ETGs for the central and satellite subsamples, panels (b) and (c), respectively. 
For the SDSS fractions $f^{c}_{E}(\ms)$ and $f^{s}_{E}(\ms)$, we perform MCMC multiparametric fits to a composition of two analytic Sigmoid functions,  following the procedure described in \cite{Rodriguez-Puebla+2013}. The final analytic function is:
\begin{equation}
f^{j}_{E}(\ms) = \frac{1-A}{1+e^{-\gamma_1(x_{C,1}+x_{0,1})}}
+ \frac{A}{1+e^{-\gamma_2(x_{C,2}-x_{0,2})}},
\label{eq:Frac_e}
\end{equation}
where $j=c$ or $s$, $x_{C,i} = \ms / \mathcal{M}_{C,i}$, with $i = 1,2$. 
For the overall fraction $f^s(\ms)$, we use an analytic function composed of a Sigmoid and constant function given by
\begin{equation}
f^{s}(\ms) = 1-\left[A\cdot\frac{1}{1+e^{-\gamma(x_{C}-x_{0})}}+H\right],
\label{eq:Frac_s}
\end{equation}
where $H$ is the constant function. Here, the Sigmoid normalization factor is defined as $A\equiv 1 - H$. 

The obtained fits are shown in Figure \ref{fig:fractions} with the solid gray lines. The fractions in the lower panels were calculated from the fractions of the upper panels. The dashed gray lines in all the panels are just the respective complementary fractions. 

Finally, in \S\S \ref{sec:DS18} we explore the effects on our results of using different morphological classification than the one used here. The alternative classification was that of \citep[][]{Dominguez-Sanchez+2018}. In Figure \ref{fig:fractions} we show  with brown colours the same fractions defined in this Appendix but using the  \citep[][]{Dominguez-Sanchez+2018} morphologies for the \xG\ and SDSS galaxies. Interestingly, now the excess in the \xG\ fraction of ETGs with respect to SDSS at masses lower than $\sim 3\times 10^{10}$ \msun\ is less than when using the \citet[][]{Huertas-Company+2011} classification, while for masses larger than this, there is now a lack  of ETGs in \xG.

\section{Procedure for reestimating the upper limits of \xG}
\label{App:upper-limits}

\subsubsection{Upper limits of ETGs}
\label{ETG-corrs}

In Paper I, based on ATLAS$^{\rm 3D}$ results, we re-scaled by distance the \texttt{GASS} upper limits of ETGs to use these valuable data along with those of ATLAS$^{\rm 3D}$ and other samples. 
%%VAR
To do so, we decreased the upper limits by $(D_{i}(z)/\overline{D}_{\rm ATLAS^{3D}})^{2}$, being $D_{i}$ the luminosity distance of each \texttt{GASS}  ETG and $\overline{D}_{\rm ATLAS^{3D}}=25$ Mpc the median luminosity distance of ATLAS$^{\rm 3D}$.
The key assumption behind this exercise is that ETGs of similar masses from \texttt{GASS} and ATLAS$^{\rm 3D}$ follow the same \mha\ statistical distribution despite their slightly different ages.   For $\sim 25\%$ of the ETGs upper limits in \texttt{GASS}, we actually assigned them a detection taking into account that in between the  \texttt{GASS} detection limit and this limit shifted to 25 Mpc, $\sim 25\%$ of galaxies in the  ATLAS$^{\rm 3D}$ sample were detected. For the remaining 75\% of \texttt{GASS} upper limits, we re-calculated them using the distance of 25 Mpc. 
That is, even for such a small distance, yet a significant fraction of \texttt{GASS} ETGs would remain as non-detected but their re-scaled upper limits to those of ATLAS$^{\rm 3D}$ result much lower than the reported ones. 
These upper limits along with those from other ETG samples compiled in Paper I, pile up around values in \RHI\ of $10^{-3}-10^{-4}$.  
The larger the mass, the smaller these values. From the performed continuous fit to the observed \RHI\ distributions in \ms\ bins, the \RHI\ values where the upper limits pile up were constrained by the function $\mathcal{R}_{1}(\ms)$, see Eq. (11) in Paper II. 
The values of $\mathcal{R}_{1}(\ms)$ correspond roughly to those where the top-hat functions start in the conditional PDFs for ETGs shown in Figure \ref{fig:logRHI-logMs-PDFs}. The fraction of galaxies in the top-hat functions correspond to the fractions of non detections.\footnote{To estimate the \RHI\ distributions of ETGs, in Paper I we assumed that the true \RHI\ values should be up to $\sim 1$ dex below the upper limit values after corrections and survival analysis, following a uniform distribution. This is why the \RHI\ conditional PDFs shown in  Figure \ref{fig:logRHI-logMs-PDFs} have a top-hat distribution of $\sim1$ dex width at their low-\RHI\ ends; see Paper I for arguments in favor of this assumption and for a discussion.}
As expected, for $\ms\gtrsim 10^{10}$ \msun, the values of $\mathcal{R}_{1}(\ms)$ are close to the upper limits of ATLAS$^{\rm 3D}$. However, have in mind that in Paper I we included other galaxy samples besides \texttt{GASS} and ATLAS$^{\rm 3D}$.

 Based on the analysis of Paper I, we proceed here as follows in order to re-estimate the \texttt{xGASS} upper limits of ETGs:

\begin{enumerate}

\item From the empirical ETG \RHI\ conditional PDFs %given \ms\
reported in Paper II, calculate the fraction of galaxies that lie in each stellar mass bin in between the \texttt{GASS} and ATLAS$^{\rm 3D}$ \RHI\ detection limits (as done in Paper I), and in between the \RHI\ detection limit of the of low-\texttt{xGASS} and $\mathcal{R}_{1}(\ms)$ (recall that in  ATLAS$^{\rm 3D}$ there are not low-mass galaxies).

\item  Assign \RHI\ values to a fraction of the \texttt{xGASS} upper limits at each \ms\ bin equal to the respective fraction as calculated in (i). To do so, pick randomly \RHI\ values from the empirical ETG \RHI\ conditional PDFs in the \RHI\ ranges determined in (i) (in Paper I, for \texttt{GASS} galaxies, a uniform distribution was assumed). 

\item For the (large) fraction of galaxies with upper limits that were not assigned an \RHI\ value, lower their upper limits by a factor $(D_i(z)/25 Mpc)^{2}$, where $D_i(z)$ is in Mpc. This is equivalent to say that these galaxies, with similar observational setups and signal-to-noise ratios as used in \texttt{xGASS} and \texttt{GASS}, will remain undetected in \HI\ at the distance of 25 Mpc, but their upper limits are re-calculated accordingly to this distance. 

\end{enumerate}

It is worth of mentioning that for $\ms>10^{10}$ \msun, the values of the fractions calculated in (i)  are around $30-40\%$, larger than the $\sim 25-30\%$ fraction of galaxies detected by ATLAS$^{\rm 3D}$ in between the detection limit of this survey and the one of \texttt{GASS} (see Paper I).

\subsubsection{Upper limits of LTGs}\label{LTG-corrs}

From Figure \ref{fig:xGASS-data-and-fracs} we see that the \xG\ detection limits lie in the very low end of our empirical \RHI\ conditional PDFs of LTGs shown in Figure \ref{fig:logRHI-logMs-PDFs}. The fraction of LTGs with upper limits that pile up around these limits is relatively small. Note that if these galaxies were closer, then they likely would have been detected in \HI, as is the case for galaxies from the closer HRS sample, see Paper I. % as the HRS sample. 
Thus, we convert the upper limit of a given LTG to a \textit{detection} with the \RHI\ value randomly picked out from the tail of the empirical \RHI\ conditional PDF given \ms\ from Paper II.

\section{Results without taking into account corrections to \xG\ }
\label{App:no-corrections}

%%%%%%%%%%%%%%%%FIGURE%%%%%%%%%%%%%%%
\begin{figure*}
	%trim=l b r t
	\includegraphics[trim = 5mm 36mm 55mm 12mm, clip, width=\textwidth]{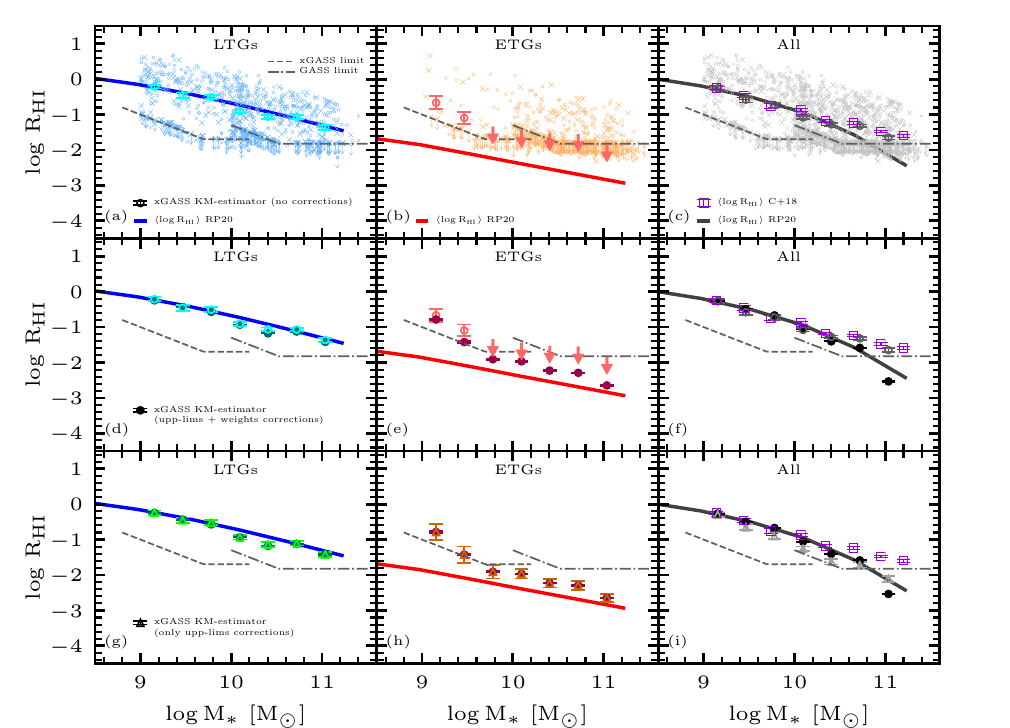} %, height=0.70\textwidth, height=180pt
	\caption{
	\textit{Upper panels:} \xG\ galaxies in the $\log\RHI-\log\ms$ diagram, as in Fig. \ref{fig:xGASS-data-and-fracs}. Large empty circles with error bars are the logarithmic means and the error of the mean in \ms\ bins obtained with the KM estimator without taking into account our procedure for the upper limits of \xG, nor the correction by morphology/environment. The solid lines show the mean \RHI--\ms\ relations from Paper II. In panel (b), means above $\ms\sim 5\times 10^9$ \msun\ are shown as arrows given that the fraction of non-detections are $>50\%$ in these mass bins (see text). In panel (c), the violet empty squares are the logarithmic means and error of the mean as reported in \citet[][]{Catinella+2018}.
	\textit{Middle panels:} Logarithmic means and their error on the mean using the KM estimator with (filled circles as in Figure \ref{fig:xGASS-correlations}) and without (empty circles or arrows, as in the upper panels) including our corrections to upper limits and morphology/environment bias of \xG. \textit{Lower panels:} Logarithmic means and their errors on the mean obtained with the KM estimator taking into account our procedure for the upper limits and weighting by morphology/environment (filled circles, as in the middle panels) and not weighting by morphology/environment (empty triangles).
	}
	\label{fig:RHI-Ms-raw} 
\end{figure*}
%%%%%%%%%%%%%% END FIGURE%%%%%%%%%%%%%

%%%%%%%%%%%%%%%%FIGURE%%%%%%%%%%%%%%%
\begin{figure*}
	%trim=l b r t
	\includegraphics[trim = 4mm 40mm 0mm 10mm, clip, width=0.95\textwidth]{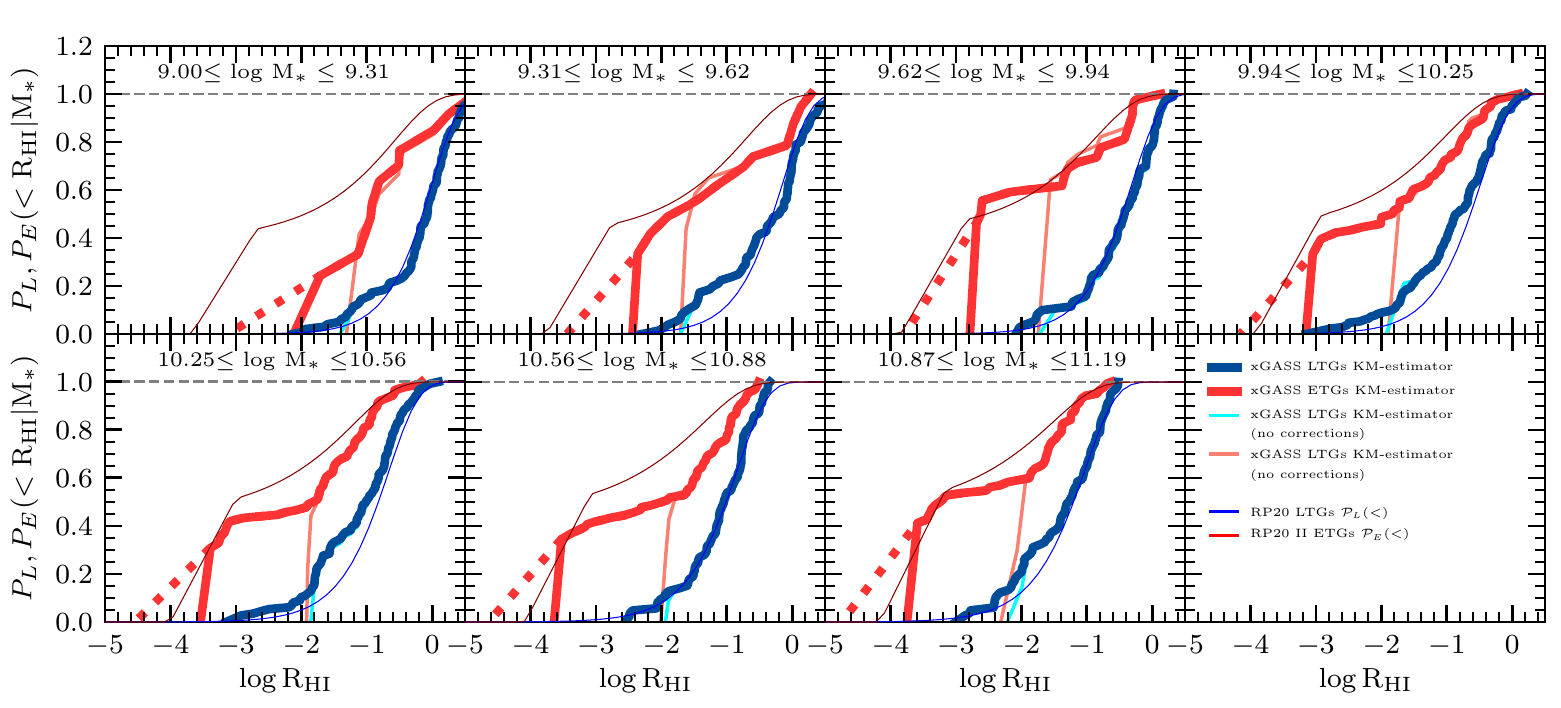} %, height=0.70\textwidth, height=180pt
	\caption{Same \RHI\ conditional CDFs from Fig. \ref{fig:xGASS-CDFs} and results without taking into into account our procedure for the upper limits of \xG\, nor the correction by morphology/environment (lighter colours).}
	\label{fig:CDFs-comparison} 
\end{figure*}
%%%%%%%%%%%%%% END FIGURE%%%%%%%%%%%%%

The upper panels of Fig. \ref{fig:RHI-Ms-raw} are as the upper panels of Fig. \ref{fig:xGASS-correlations} but without taking into account our procedure for the upper limits of \xG, nor the correction by morphology/environment (the respective data are presented in tabulated form in the Supplementary Material). Here, instead of the standard deviation, we plot the error of the mean. For ETGs,  in the stellar mass bins above $\ms\sim 5\times 10^9$ \msun\ the obtained means with the KM estimator are shown with an arrow. This is because the fraction of non-detections are higher than 50\% in these mass bins, in which cases the KM estimator provides uncertain results; the means should be taken as upper bounds while the error on the means (or standard deviations) are meaningless. 
In the middle panels, we compare the \RHI\ means and errors on the mean obtained with the KM estimator with and without including our corrections to upper limits and morphology/environment bias of \xG. For LTGs, the results are almost indistinguishable from each other. 
For ETGs less massive than $\ms\sim 5\times 10^9$ \msun, the means are slightly higher when our procedure for the upper limits is not taken into account. For $\ms\gtrsim 5\times 10^9$ \msun, the means are only an estimate of the upper bound.  For the whole sample, combining LTGs and ETGs, the KM results without taking into account the procedure for upper limits are only slightly below to those reported in \citet[][]{Catinella+2018}, who assigned \RHI\ values to non-detections equal to their upper limit values. 
Finally, in the lower panels of  Fig. \ref{fig:RHI-Ms-raw}, we compare the \RHI\ means obtained with the KM estimator taking into account our procedure for the upper limits but applying and not applying the weights by morphology/environment (the respective data are presented in tabulated form in the Supplementary Material). The weights (mainly by morphology) slightly increase the mean \RHI\ values for masses below $\ms\sim 5\times 10^{10}$ \msun, while for the highest masses, the weights decrease \RHI\ by $\sim 0.3$ dex.

In Fig. \ref{fig:CDFs-comparison} we reproduce the \RHI\ conditional CDFs plotted in Fig. \ref{fig:xGASS-CDFs} and compare them with those without taking into into account our procedure for the upper limits of \xG. For LTGs, the CDFs in both cases are very similar, excepting at the low-\RHI\ end in the most massive bins. For ETGs, when the procedure for the upper limits is not taken into account, the CDFs undergo a sharp cut at relatively high values of \RHI. In this case, we can not constrain any reliable \RHI\ conditional CDF.

\section{Conservation equations}\label{app:eqs-consv}

As discussed in \S\S \ref{sec:corrections}, performing fits to xGASS CDFs must obey the law of total probability. Here, we present the ``probability conservation equations'' in order to satisfy such requirement for the whole set of galaxies, different subsets of LTGs/ETGs, centrals/satellites, and their combinations.

First, to describe the \HI\ conditional CDFs of all LTGs and ETGs, and central LTGs and ETGs (four sets of CDFs) we propose the analytic incomplete gamma function given by Eq.(\ref{eq:gamma-tot}) for each one of these populations.

The remaining five sets of \HI\ CDFs to be used also for the fitting procedure are described by the below listed five equations that obey the law of total probability, and that allow us to use the above mentioned four sets of CDFs for calculating these five sets of CDFs. Such equations require information on different fractions of populations and subpopulations of galaxies as a function of \ms. In Appendix \ref{app:fractions} we discuss how we estimate these fractions. For simplicity, we do not show the dependence of these fractions on \ms\ in the following equations:

\begin{itemize}
    \item \textit{HI CDFs of the whole sample:}
    \begin{equation}\label{eq:cons-tot}
        P^{T}(<\RHI|\ms)=f^{L}\cdot P^{L}(<\RHI|\ms)+f^{E}\cdot P^{E}(<\RHI|\ms)
    \end{equation}
    where $f^{E}$ and $f^{L}$ are the fractions of ETGs and LTGs, respectively; $f^{E}+ f^{L}=1$.
    
    \item \textit{HI CDFs of the subsample of centrals:}
    \begin{equation}\label{eq:cons-cen}
        P^{c}(<\RHI|\ms)=f^{c}_{L}\cdot P^{c}_{L}(<\RHI|\ms)+f^{c}_{E}\cdot{P}^{c}_{E}(<\RHI|\ms)
    \end{equation}
    where $f^{c}_{E}$ and $f^{c}_{L}$ are the fractions of centrals that are ETGs and LTGs, respectively; $f^{c}_{E}+f^{c}_{L}=f^c$
    
    \item \textit{HI CDFs of the subsample of satellites}
    \begin{equation}\label{eq:cons-sat}
        P^{s}(<\RHI|\ms)=\frac{1}{f^{s}}\left[P^{T}(<\RHI|\ms)-f^{c}\cdot P^{c}(<\RHI|\ms)\right]
    \end{equation}
    where $P^{T}(<\RHI|\ms)$ and $P^{c}(<\RHI|\ms)$ are the total and centrals CDFs given by eqs. (\ref{eq:cons-tot}) and (\ref{eq:cons-cen}) respectively. $f^{c}$ and $f^{s}$ are the fraction of centrals and satellites, $f^{c}+f^{s}=1$.

    \item \textit{HI CDFs of the subsample of satellites that are LTGs}
    \begin{equation}\label{eq:cons-sat-LTGs}
        P^{s}_{L}(<\RHI|\ms)=\frac{1}{f^{L}_{s}}\left[P_{L}^{T}(<\RHI|\ms)-f^{L}_{c}\cdot P^{c}_{L}(<\RHI|\ms)\right]
    \end{equation}
    where $P_{L}^{T}(<\RHI|\ms)$ and $P^{c}_{L}(<\RHI|\ms)$ are the total LTGs and LTGs centrals CDFs analytic fits given by eq.(\ref{eq:gamma-tot}) respectively. $f^{L}_{c}$ and $f^{L}_{s}$ are the fractions of LTGs that are centrals and satellites, $f^{L}_{c} + f^{L}_{s}= f^L$

    \item \textit{HI CDFs of the subsample of satellites that are ETGs}
    \begin{equation}\label{eq:cons-sat-ETGs}
        P^{s}_{E}(<\RHI|\ms)=\frac{1}{f^{E}_{s}}\left[P_{E}^{T}(<\RHI|\ms)-f^{E}_{c}\cdot P^{c}_{E}(<\RHI|\ms)\right]
    \end{equation}
    where $P_{E}^{T}(<\RHI|\ms)$ and $P^{c}_{E}(<\RHI|\ms)$ are the ETGs and ETG centrals CDFs analytic fits given by eq.(\ref{eq:gamma-tot}), respectively. $f^{E}_{c}$ and $f^{E}_{s}$ are the fractions of ETGs that are centrals and satellites, $f^{E}_{c}+f^{E}_{s}= f^E$.
    \end{itemize}

%%%%%%%%%%%%%%%%%%%%%%%%%%%%%%%%%%%%%%%%%%%%%%%%%%

% Don't change these lines
\bsp	% typesetting comment
\label{lastpage}
\end{document}